\documentclass[10pt,twocolumn]{IEEEtran}
\usepackage{tabularx}
\usepackage[T1]{fontenc}
\usepackage{algorithmic}
\usepackage{graphicx}
\usepackage{comment}
\usepackage{amsmath,amssymb,amsfonts}
\usepackage{array}
\usepackage{textcomp}
\usepackage[font=small]{caption}
\usepackage{paralist}
\usepackage{lettrine}
\usepackage{amsthm}
\usepackage{cite}
\usepackage{xcolor}
\usepackage{cancel}
\usepackage{booktabs}
\usepackage{subfig}
\usepackage{stfloats}
\usepackage{url}
\usepackage{lipsum}
\usepackage{enumitem}
\usepackage{mathtools}
\usepackage{cuted}
\usepackage{svg}
\usepackage{makecell}
\usepackage{acronym}
\usepackage{multicol}
\usepackage{multirow}
\usepackage[ruled,vlined]{algorithm2e}
\usepackage{authblk}
\usepackage{fixltx2e}
\usepackage{tikz}
\usepackage{hyperref}

\usepackage[utf8]{inputenc}
\usepackage{cite}
\usepackage{floatrow}

\usepackage{amsmath,amssymb,amsfonts,stfloats}
\usepackage{tabularx}
\usepackage{algorithmic}
\usepackage{graphicx}
\usepackage{textcomp}
\usepackage{floatrow}
\usepackage{xcolor}
\usepackage{booktabs}

\usepackage{multicol}
\usepackage{acronym}
\usepackage{url}
\usepackage{mathtools}

\begin{document}

\title{Integrating Phase-Coherent Multistatic Imaging in Downlink D-MIMO Networks 
}

\author{Dario~Tagliaferri,~\IEEEmembership{Member,~IEEE}, Silvia~Mura,~\IEEEmembership{Member,~IEEE}, Musa~Furkan~Keskin,~\IEEEmembership{Member,~IEEE}, Sauradeep~Dey, Henk~Wymeersch,~\IEEEmembership{Fellow,~IEEE}
\thanks{Dario~Tagliaferri and Silvia~Mura are with the Department of Electronics, Information and Bioengineering (DEIB), Politecnico di Milano, Milano, Italy (emails: \{dario.tagliaferri,silvia.mura\}@polimi.it).}
\thanks{Musa~Furkan~Keskin, Sauradeep~Dey and Henk~Wymeersch are with the Department of Electrical Engineering, Chalmers University of Technology, Gothenburg, Sweden (emails: \{furkan,deysa,henkw\}@chalmers.se).}
\thanks{This work is partially supported by the SNS JU project 6G-DISAC under the EU’s Horizon Europe research and innovation programme under Grant Agreement No 101139130, and the Swedish Research Council (VR) through the project 6G-PERCEF under Grant 2024-04390.}
}

\maketitle

\begin{abstract}
This paper addresses the challenge of integrating multistatic coherent imaging functionalities in the downlink (DL) of a phase-coherent distributed multiple input multiple output (D-MIMO) communication network. During DL, the D-MIMO access points (APs) jointly precode the transmitted signals to maximize the spectral efficiency (SE) at the users (UEs) locations. However, imaging requires that \textit{(i)} a fraction of the APs work as receivers for sensing and \textit{(ii)} the transmitting APs emit AP-specific and orthogonal signals to illuminate the area to be imaged and allow multistatic operation.   
In these settings, our contribution is twofold. We propose a novel distributed integrated sensing and communication (D-ISAC) system that superposes a purposely designed AP-specific signal for imaging to the legacy UE-specific communication one, with a tunable trade-off factor. We detail both the imaging waveform design according to the \textit{extended orthogonality condition} and the space-frequency precoder design. Then, we propose an optimized selection strategy for the receiving APs, in order to maximize imaging performance under half-duplex constraints. Extensive numerical results prove the feasibility and benefits of our proposal, materializing the potential of joint multistatic imaging and communications in practical D-MIMO deployments.
\end{abstract}

\begin{IEEEkeywords}
Phase-coherent multistatic imaging, downlink, D-MIMO, D-ISAC, trade-off.
\end{IEEEkeywords}

\section{Introduction}

Integrated Sensing and Communication (ISAC) is a key pillar of next generation wireless systems, merging radar sensing and communication into a unified system with shared hardware and radio resources, for various applications~\cite{Nuria_Proc_IEEE_2024}. 
Early research focused on single-terminal ISAC setups, either \textit{monostatic} (co-located transmitter and sensing receiver \cite{Keskin2025_holistic_monostatic}) or \textit{bistatic} (separated \cite{Giorgetti_bistatic_2024}), balancing data transmission and environmental sensing. This led to the design of joint waveforms with tunable trade-offs across spatial \cite{liu2018toward} and time-frequency \cite{Tagliaferri_DD_2024,Wymeersch2021} domains. 
More recent research has expanded ISAC from single devices to networks, where multiple nodes cooperate to improve sensing accuracy, capacity, and reliability. On the communication side, distributed multiple-input-multiple output (D-MIMO) in cell-free architectures has shown definite advantages over massive MIMO~\cite{Buzzi_cell-free}. For sensing, deploying spatially distributed nodes enables \textit{multistatic} ISAC---combining monostatic and bistatic measurements---which provides extra spatial diversity, improving resolution, localization accuracy, and robustness~\cite{chernyak1998fundamentals}.


The literature on distributed and cooperative ISAC (D-ISAC) networks is still recent and mostly focused on orthogonal frequency division multiplexing (OFDM), the legacy waveform adopted by the third generation partnership project (3GPP). A high-level overview of D-ISAC benefits and challenges is given in \cite{Masouros2024_cellfreeISAC_magazine}, while interference management through coordinated beamforming across multiple monostatic nodes is addressed in \cite{Masouros2024_networked_ISAC_interference}. An optimal antenna allocation strategy balancing localization accuracy and communication constraints is proposed in \cite{Masouros_antenna_topology_DISAC}. The joint beamforming design problem for D-ISAC has been investigated in some recent works, for instance \cite{Alkhateeb_cellfreeISAC,Li_BF_Design_grouping_CFISAc,Lou_BF_design_CFISAC,Masouros2025_distributedISAC}, where the Cramér–Rao lower bound (CRLB) on target localization is exploited for communication–sensing beamforming trade-offs. In particular, \cite{Alkhateeb_cellfreeISAC} considers fully digital nodes with separate RF chains for communication and sensing, whereas \cite{Masouros2025_distributedISAC} proposes a weighted superposition of beamformers. Further CRLB-based insights are provided in \cite{Giorgetti2024_cooperativesensing}. 
Other contributions, such as \cite{11027423}, address waveform design to ensure inter-node orthogonality, enhancing sensing performance similar to MIMO radar. For a recent survey of D-ISAC research, see \cite{galappaththige2025cellfreeintegratedsensingcommunication}.

The existing D-ISAC works share two main features: \textit{(i)} they integrate \textit{target localization} within D-MIMO or cell-free networks, typically under a communication-centric perspective \cite{Alkhateeb_cellfreeISAC,Li_BF_Design_grouping_CFISAc,Lou_BF_design_CFISAC,Masouros2025_distributedISAC}; \textit{(ii)} D-ISAC nodes (or access points (APs) in cell-free terminology) perform incoherent sensing, i.e., fusing measurements from monostatic/bistatic pairs without exploiting carrier phase information \cite{Giorgetti2024_cooperativesensing}. Most prior D-ISAC works focus on \textit{state estimation} (position, velocity, orientation) of desired targets, typically by minimizing CRLB-based expressions under communication constraints. However, CRLB is tractable, i.e., it exists in closed form, only for a few well-separated targets, as the Fisher information matrix (FIM) grows quadratically with the number of closely spaced targets, where the effect of waveform sidelobes degrades estimation performance \cite{win_Localization}. In scenarios with an arbitrary number of close targets, \textit{imaging} is more appropriate. Imaging reconstructs a map of the \textit{complex reflectivity} of the environment at a given carrier frequency, from which the number, location, shape, and size of targets can be inferred. Its key performance metric is the \textit{effective resolution}, i.e., the capability to separate closely spaced targets given D-ISAC network configuration, bandwidth, and waveform. Imaging is central to remote sensing \cite{Moreira_SAR_tutorial}, but has received limited attention in ISAC, and mostly for single nodes \cite{Li2024,11087660,BelliniMultiView,FanLiu_imaging,ManzoniCOSMIC}. For example, \cite{Li2024} studies a monostatic full-duplex BS serving UEs while imaging its surroundings via joint beamforming, pointing at fictitious ''virtual UEs''; the work \cite{11087660} formulates imaging as an inverse problem to estimate voxel-based reflectivity magnitudes while preserving communication reliability. Other works investigate metasurfaces \cite{BelliniMultiView}, satellite-based imaging \cite{FanLiu_imaging}, or orthogonal waveform design for single-node ISAC~\cite{ManzoniCOSMIC}.

A crucial distinction in D-ISAC lies in \textit{phase-incoherent} and phase-\textit{phase-coherent} systems. Phase-incoherent sensing enhances localization accuracy---improving signal-to-noise ratio (SNR) and the geometric dilution of precision (GDOP)---compared to single node ISAC \cite{Giorgetti2024_cooperativesensing}, but brings little to no gain in imaging resolution, which remains limited by the bandwidth and number of antennas of individual nodes. In contrast, phase-coherent systems directly exploit the carrier phase, enabling distributed coherent arrays with higher imaging performance. While closed-form expressions exist for the \textit{theoretical (Rayleigh) resolution}~\cite{resAnalysis_2024}, determined by spatial aperture and involving only the main lobe of the \textit{spatial ambiguity function} (SAF), the effective resolution is strongly topology-dependent due to SAF sidelobes, grounding phase-coherent imaging in diffraction theory~\cite{manzoni2024wavefield}. Phase-coherent AP operation has been extensively studied in cell-free D-MIMO \cite{demir2021foundations}, and most of the (few) works have addressed its use in distributed radar networks (D-RNs) \cite{bolomey1990microwave,Waldschmidt2024,tagliaferri2024cooperative}, without integration of communication functionalities.
Research on coherent D-ISAC systems is still in its early stages \cite{10097213,10540249,IIAC_3D_imaging,zhi2025nearfieldintegratedimagingcommunication}. Works such as \cite{10097213,10540249,IIAC_3D_imaging} formulate imaging as an inverse problem, where communication signals are used to reconstruct a 2D/3D occupancy map of the environment over a voxel grid, using one or more APs and UEs. To avoid artifacts, the number of voxels is typically constrained by the system’s spatial degrees of freedom (total antennas at UEs and APs). Higher resolution can be pursued by leveraging \textit{a-priori information} on targets (e.g., shape) or by imposing \textit{sparsity constraints} as in tomographic imaging to regularize the inversion~\cite{Wang_RTI}. The latter assumptions may limit the application in complex environments, where numerous targets of different shape can be present. The most recent work \cite{zhi2025nearfieldintegratedimagingcommunication} extends this research line to 3D near-field imaging of non-isotropic targets, again via inverse reflectivity reconstruction with multiple nodes. However, none of the existing D-ISAC works \textit{jointly} address phase-coherent communication and imaging. Previous studies addressed phase-coherent communication \cite{demir2021foundations} or imaging \cite{Waldschmidt2024,bolomey1990microwave,tagliaferri2024cooperative,10097213,10540249,IIAC_3D_imaging,zhi2025nearfieldintegratedimagingcommunication}, without exploring the inherent trade-offs of mutual integration into a D-MIMO system.

\vspace{-0.3cm}\subsection{Problem Statement and Contributions}

In this paper, we address the trade-offs between communication and imaging in D-MIMO systems. We propose a downlink (DL) phase-coherent D-ISAC system, where APs simultaneously serve multiple UEs while performing high-resolution multistatic phase-coherent imaging of a region of interest (ROI). Three key challenges are addressed: \textit{(i)} D-MIMO in time-division duplexing in DL uses \textit{UE-specific} precoded information signals, leveraging uplink (UL) channel estimates and exploiting DL/UL channel reciprocity~\cite{7880691}, whereas imaging requires \textit{AP-specific} orthogonal signaling; \textit{(ii)} spatial focusing for communication does not cover large areas in a single snapshot; \textit{(iii)} imaging requires some APs to act as Rx in typically half-duplex networks, while D-MIMO works with Tx-only APs in DL. Our approach introduces a novel waveform, spatial precoding, and AP selection strategy to jointly enable communication and phase-coherent multistatic imaging.
The main contributions of the paper are as follows:
\begin{enumerate}

    \item \textbf{Space-time-frequency signal design}. We propose a novel system design that superposes, in frequency and time (FT), the UE-specific communication signal (OFDM), \textit{coherently} focused at the UE location, to an AP-specific waveform for imaging, designed in the delay and Doppler (DD) domain and spatially precoded to \textit{incoherently} illuminate the ROI. The D-ISAC performance is dictated by a trade-off factor that regulates the power scaling between communication and sensing waveforms in frequency, time and space. The imaging signal is designed according to the \textit{extended orthogonality condition}. Therefore, we first extend the waveform design in \cite{ManzoniCOSMIC} to improve imaging in multistatic D-ISAC networks that do not adopt the extended orthogonality condition. Then, we detail the mixed coherent-incoherent space-frequency precoding scheme for D-ISAC, that leverages a fictitious LOS channel between the Tx APs and the ROI. 

    \item \textbf{AP selection strategy}. We propose a new selection strategy to group half-duplex APs into disjoint Tx and Rx sets, according to the constrained minimization of the entropy of the multistatic coherent SAF. In the optimization, we enforce the half-duplex constraint and a maximum number of Rx APs, in order to limit the SE reduction during the D-ISAC DL operation. Our method allows maximizing sensing performance, especially for communication-centric D-ISAC systems.

    \item \textbf{Insights and practical guidelines}. We analyze the performance of the proposed D-ISAC system in terms of spectral efficiency (SE), sensing signal-to-interference-plus-noise ratio (SINR) and SAF entropy, varying system parameters, against D-MIMO and D-RN benchmarks. We derive useful insights and design guidelines for practical implementation, proving the benefits of our proposed D-ISAC system. In particular, we show that \textit{(i)} dense ISAC networks made of many nodes with few antennas are preferable to sparse networks with massive nodes; \textit{(ii)} for a fixed number of APs, the larger the ROI the less is the number of possible orthogonal signals for the APs, posing a further trade-off between imaging quality and ROI size; \textit{(iii)} in communication-centric D-ISAC networks, an optimal Rx AP selection yields significant benefits for imaging. 

\end{enumerate}

We highlight the fundamental differences between our work and the existing D-ISAC literature~\cite{bolomey1990microwave,10097213,10540249,IIAC_3D_imaging,zhi2025nearfieldintegratedimagingcommunication}. While the latter focuses on optimizing the reconstruction of the target’s shape \textit{given} the communication system (i.e., with OFDM communication signals), and whose performance are implicitly environment-specific, our goal is to judiciously design the D-ISAC system from bottom-up (and study the related trade-offs) that lead to the maximization of imaging performance, \textit{prior} to the application of any inverse reconstruction technique. Therefore, our work is agnostic to the type of targets, and focuses on system design. In this sense, works~\cite{bolomey1990microwave,10097213,10540249,IIAC_3D_imaging,zhi2025nearfieldintegratedimagingcommunication} constitute a subsequent processing step w.r.t. our proposal.

\begin{figure}[!t]
    \centering
    \includegraphics[width=\linewidth]{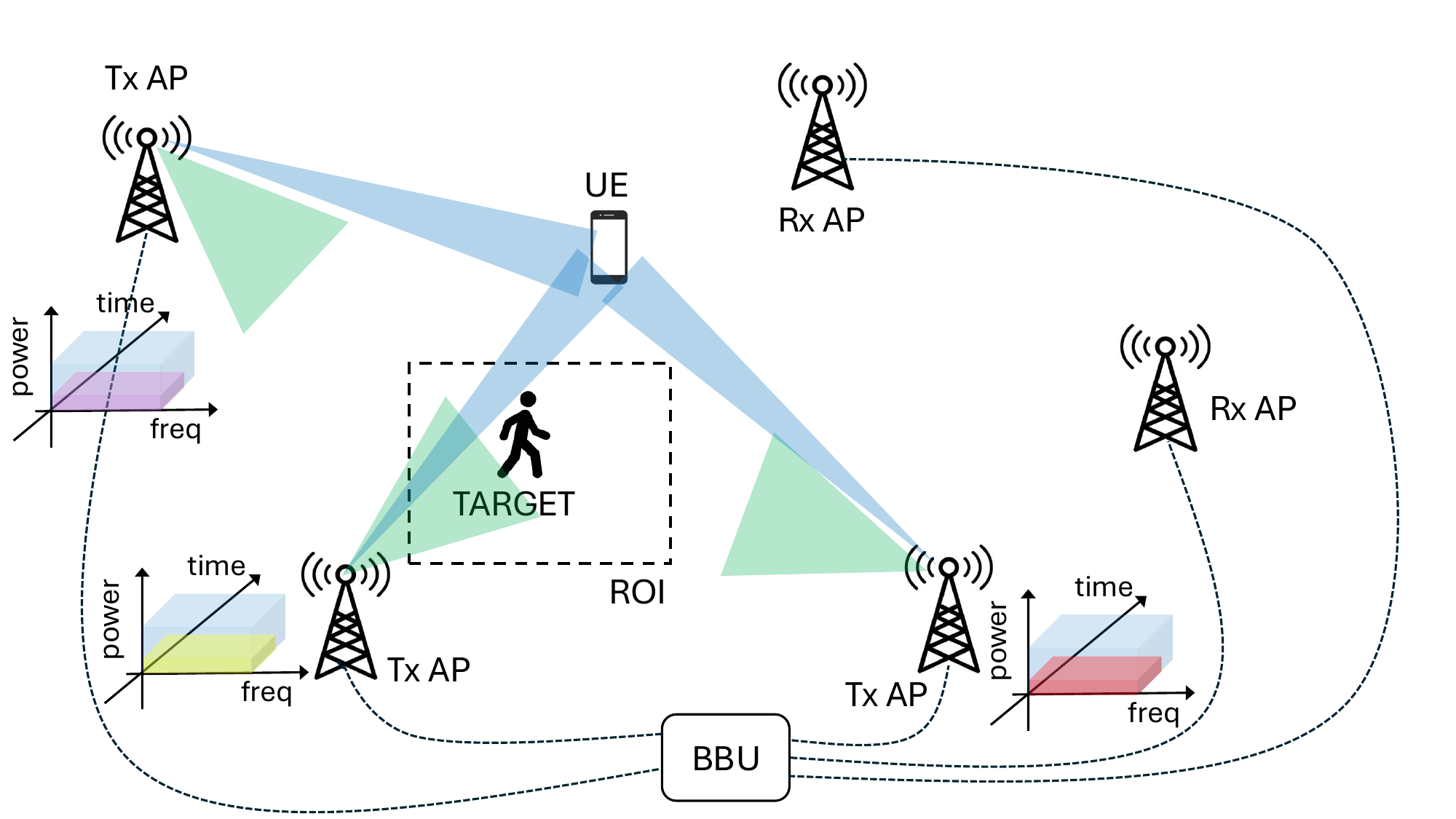}
    \caption{Illustration of the proposed D-ISAC system: A subset of the APs (Tx APs) emit the superposition of a UE-specific communication signal and an AP-specific sensing signal. The latter is received by a set of Rx APs (whose set is disjoint from the Tx APs one), that generate the image of the targets in the ROI. While the communication signal is spatially precoded to coherently focus in the UE position, the sensing signals are precoded to illuminate the ROI. }
    \label{fig:intro}
\end{figure}

\section{System Model}\label{sec:system_model}

\subsection{Scenario}\label{subsec:scenario}

We consider the D-ISAC system in Fig. \ref{fig:intro}, composed of $N$ APs aiming at jointly serving $Q$ UEs during the DL, while generating a phase-coherent multistatic image of the targets in a ROI at the highest possible resolution. The frequency of operation is $f_0$ and the employed bandwidth is $B$. The geometry of the system is arbitrary. The $n$-th AP, $n=1,...,N$, is located in $\mathbf{p}_n\in\mathbb{R}^{2 \times 1}$, oriented in space by angle $\psi_n$ and equipped with a uniform linear array (ULA) with $L$ antennas, spaced by $d=\lambda_0/2$ ($\lambda_0$ is the carrier wavelength). The APs are equipped with a single array and cannot work simultaneously as Tx and Rx, operating in \textit{half-duplex}. The $q$-th single-antenna UE, $q=1,...,Q$, is located in $\mathbf{x}_q\in\mathbb{R}^{2 \times 1}$. We have $Q \leq NL$, as typical in D-MIMO or cell-free networks, where all the APs share the same base band unit (BBU). The ROI is centered in $\overline{\mathbf{x}}$, of size $\Delta_x \times \Delta_y$, and contains $U$ targets, each located in $\mathbf{x}_u \in\mathbb{R}^{2 \times 1}$, $u=1,...,U$. To enable this novel D-ISAC system, some APs are selected to operate as Rx, while the remaining ones as Tx. We denote Tx and Rx sets of APs with $\mathcal{N}_{\rm tx}$ and $\mathcal{N}_{\rm rx}$, respectively, whose number are $N_{\rm tx}$ and $N_{\rm rx}$. The half-duplex condition imposes $\mathcal{N}_{\rm tx} \cap \mathcal{N}_{\rm rx} = \emptyset$ and $N_{\rm tx} + N_{\rm rx} = N$. 

\vspace{-0.3cm}\subsection{Transmitted ISAC Signal}\label{subsec:Txsignal}

The Tx signal by each AP uses an OFDM waveform. We consider a FT resource grid of $M$ subcarriers and $K$ OFDM symbols, defined as
\begin{equation}\label{eq:TFgrid}
    \begin{split}
        \Lambda = \bigg\{ m\Delta f, kT \,\big|\, m \in \left[-\frac{M}{2},\frac{M}{2}\hspace{-0.1cm}-\hspace{-0.1cm}1\right],k \in [0,K\hspace{-0.1cm}-\hspace{-0.1cm}1] \bigg\},
    \end{split}
\end{equation}
where $\Delta f$ is the subcarrier spacing and $T$ is the OFDM symbol duration, comprising the cyclic prefix. The occupied bandwidth is $B = M\Delta f$, and the total duration of the DL burst is $KT$. For completeness, we also define the resource grid in the dual DD domain as
\begin{equation}\label{eq:DDgrid}
    \begin{split}
        \widetilde{\Lambda} = \bigg\{ \ell \Delta \tau, p \Delta \nu \,\big|\, &\ell \in [0,M\hspace{-0.1cm}-\hspace{-0.1cm}1], p \in\left[-\frac{K}{2},\frac{K}{2}\hspace{-0.1cm}-\hspace{-0.1cm}1\right] \bigg\},
    \end{split}
\end{equation}
where $\Delta \tau = 1/B$ is the delay resolution of the system, while $\Delta \nu = 1/(KT)$ is the Doppler resolution. The delay dimension is sometimes called \textit{fast-time}, while OFDM symbols are usually identified as \textit{slow-time} for sensing. Notice that the number of samples along frequency (time) may not be necessarily equal to the number of samples along delay (Doppler). The FT and DD domains are inter-related by a pair of direct and inverse Fourier transforms, as detailed in \cite{1088793}. 

To enable the integration of phase-coherent sensing into the legacy OFDM FT grid $\Lambda$, the $n$-th Tx AP, $n\in \mathcal{N}_{\rm tx}$, emits the weighted superposition of the communication symbols towards the $Q$ UEs, $X_{q,\mathrm{com}}$ that is \textit{UE-specific}, and an \textit{AP-specific} sensing signal $X_{n,\mathrm{sen}}$, as:
\begin{equation}\label{eq:TX_signal_FT}
\begin{split}
       \mathbf{s}_{n}[m,k] &= \sqrt{P} \eta \sum_{q=1}^Q \mathbf{v}^\mathrm{com}_{n,q}[m] X_{q,\mathrm{com}}[m,k] \\ & + \sqrt{P} (1-\eta) \mathbf{v}^\mathrm{sen}_{n}[m] X_{n,\mathrm{sen}}[m,k]
\end{split}
\end{equation}
where $\mathbf{v}^\mathrm{com}_{n,q}[m] \in \mathbb{C}^{L\times 1}$ is the frequency-varying (wideband) spatial precoder for the $q$-th communication signal, $\mathbf{v}^\mathrm{sen}_{n}[m] \in \mathbb{C}^{L\times 1}$ is the precoder for the sensing signal and $\eta\in [0,1]$ is a scaling factor expressing the performance trade-off between communication and sensing.  Each element of $X_{q,\mathrm{com}}[m,k]$ is drawn from a QAM constellation with unitary power. Both communication and sensing precoders are normalized such that the total emitted power by the D-ISAC network (i.e., by the $N_{\rm tx}$ APs) on each subcarrier is $P$, as detailed in Section~\ref{subsect:precoding}.

The D-ISAC system must therefore be designed to fulfil the following needs: \textit{(i)} the effect of the sensing signal at the UEs shall be minimal, in order to not affect the capacity; \textit{(ii)} the effect of the communication signals at the sensing receivers (Rx APs) shall be low enough to have sufficient SINR to enable imaging of the targets in the ROI and \textit{(iii)} the sensing signals emitted at different Tx APs should be orthogonal, to leverage multiple bistatic AP pairs to generate the coherent image of the ROI. Apparently, the first two aspects lead to contrasting needs, since the first would push for $\eta \rightarrow 1$ (pure D-MIMO communication system) while the second for $\eta \rightarrow 0$ (pure sensing system). The proposed solution leverages the DD domain to design a set of AP-specific sensing signals that \textit{(i)} are mutually orthogonal over the DD domain and \textit{(ii)} exhibit good cross-correlation properties with the communication signals. 

\vspace{-0.4cm}\subsection{Communication Channel and Received Signal at the UEs }\label{subsec:comm_channel_RX_signal}

The communication channel from the $n$-th Tx AP to the $q$-th UE on the $m$-th subcarrier is customarily modeled as a multipath with $C$ scattering clusters 
\begin{equation}\label{eq:comm_channel}
\begin{split}
       \mathbf{h}_{n,q}[m] & = \sum_{c=1}^C \alpha_{nq,c} \mathbf{a}(\theta_{nq,c}) e^{-j 2 \pi (f_0 +  m \Delta f )\tau_{nq,c}} 
\end{split}
\end{equation}
where: \textit{(i)} $\alpha_{nq,c}$ is the complex scattering amplitude of the $c$-th path, \textit{(ii)} $\mathbf{a}(\theta_{nq,c})$ is the conventional far-field steering vector of the AP, function of the path angle $\theta_{nq,c}$ as $[\mathbf{a}(\theta_{nq,c})]_u = e^{- j \pi u \sin \theta_{nq,c}}$, \textit{(iii)} $\tau_{nq,c}$ is the delay of the $c$-th path. In \eqref{eq:comm_channel}, $c=0$ identifies the LOS path. The number of scattering clusters, $C$, is constant for all AP-UE pairs for simplicity. With $C=1$, we have a purely LOS channel. In all the following, we assume that the APs know the communication channels to the UEs. The case of imperfect channel reporting from the UEs to the APs will degrade communication performance, but it does not affect the conveyance of the concepts of the paper, and it is left for a future work. 

The model of the Rx signal at the $Q$ UEs and at the $N_{\rm rx}$ Rx APs define the performance of the ISAC system.  

The Rx signal at the $q$-th UE, due to the transmission of the $N_{\rm tx}$ Tx APs, is:
\begin{equation}\label{eq:RX_signal_UE_FT}
\begin{split}
    Y_q[m,k] 
    & = \underbrace{\sqrt{P}\eta \hspace{-0.1cm}\sum_{n \in \mathcal{N}_{\rm tx}} \hspace{-0.1cm}\mathbf{h}^{\mathsf{T}}_{n,q}[m]\mathbf{v}^\mathrm{com}_{n,q}[m] X_{q,\mathrm{com}}[m,k] }_{\text{Desired signal}\; \mathsf{S}_{q,\mathrm{com}}[m,k]}\\
    & + \underbrace{\sqrt{P}\eta \hspace{-0.1cm}\sum_{n \in \mathcal{N}_{\rm tx}} \hspace{-0.1cm}\mathbf{h}^{\mathsf{T}}_{n,q}[m] \sum_{q'\neq q} \mathbf{v}^\mathrm{com}_{n,q'}[m] X_{q',\mathrm{com}}[m,k] }_{\text{Communication Interference} \; \mathsf{INT}_{q,\mathrm{com}}[m,k]}\\
    & + \underbrace{\sqrt{P}(1-\eta) \hspace{-0.1cm}\sum_{n \in \mathcal{N}_{\rm tx}} \hspace{-0.1cm}\mathbf{h}^{\mathsf{T}}_{n,q}[m] \mathbf{v}^\mathrm{sen}_{n}[m] X_{n,\mathrm{sen}}[m,k]}_{\text{Sensing interference}\; \mathsf{INT}_{q,\mathrm{sen}}[m,k]} \\
    &+ W_{q}[m,k]
\end{split}
\end{equation}
The signal is composed of four contributions. The first ($\mathsf{S}_{q,\mathrm{com}}[m,k]$) is due to the desired signal intended to the $q$-th UE, $X_q[m,k]$. The second term ($\mathsf{INT}_{q,\mathrm{com}}[m,k]$) is the aggregate mutual user interference (MUI) from the data streams belonging to other $Q-1$ UEs. The third term ($\mathsf{INT}_{q,\mathrm{sen}}[m,k]$) is the interference from the sensing signals emitted by the Tx APs, at the $q$-th UE location. The last term, $W_{q}[m,k]$, is the additive Gaussian noise corrupting the Rx signal, $W_{q}[m,k] \sim \mathcal{CN}(0, \sigma_w^2)$. While the communication interference can be handled by the design of the precoders $\mathbf{v}^\mathrm{com}_{n,q}[m]$, $q=1,...,Q$, (once knowing the UE channels), the sensing disturbance is properly handled by the parameter $\eta$ and the design of the precoder $\mathbf{v}^\mathrm{sen}_{n}[m]$. In principle, by knowing the sensing signals at the UE side $\{X_{n,\rm sen}[m,k]\}$ $n \in \mathcal{N}_{\rm tx}$, is it possible to cancel it \cite{Quek_SIC}. However, this would add a non-negligible complexity to the legacy UE processing chain, with consequent energy consumption. Our goal, instead, is to design a D-ISAC system at the network side only, possibly tolerating a slight capacity reduction without adding further complexity for the UEs.

\vspace{-0.3cm}\subsection{Sensing Channel and Received Signal at the Rx APs }\label{subsec:sensing_channel_RX_signal}

In the general case, the sensing channel from the $n$- th Tx AP to the $r$-th Rx AP is modeled as the summation of $U$ targets within the ROI, as in \eqref{eq:sensing_channel},
\begin{equation}\label{eq:sensing_channel}
     \mathbf{H}_{n,r}[m] \hspace{-0.1cm}= \hspace{-0.1cm} \sum_{u=1}^U \beta_{nr,u} \mathbf{a}(\theta_{n,u}) \mathbf{a}^{\mathsf{H}}(\theta_{r,u}) e^{-j 2 \pi (f_0 + m \Delta f) \tau_{nr,u}} 
\end{equation}
where: \textit{(i)} $\beta_{nr,q}$ is the complex scattering amplitude, proportional to the radar cross section (RCS) of the $u$-th target. The phase of the scattering amplitude is assumed to be constant w.r.t. the specific $nr$-th AP pair, to enable phase-coherent sensing; \textit{(ii)} $\theta_{n,u}$ and $\theta_{r,u}$ are the angles of departure/arrival (AOD/AOA) to/from $\mathbf{x}_u$, respectively, \textit{(iii)} $\tau_{nr,u}$ is the two-way delay to/from the centers of the APs, 
\begin{equation}\label{eq:delay}
\tau_{nr,u} = \frac{\| \mathbf{x}_u - \mathbf{p}_n\|}{v} + \frac{\| \mathbf{p}_r - \mathbf{x}_u\|}{v}.
\end{equation}
where $v$ is the speed of light.
For the purpose of the paper, we also define the sensing channel in DD domain, reported in~\eqref{eq:sensing_channel_DD}. 
\begin{figure*}
\begin{equation}\label{eq:sensing_channel_DD}
\begin{split}    
\widetilde{\mathbf{H}}_{n,r}[\ell] & = \Delta f \sum_{m}  \mathbf{H}_{n,r}[m] \, e^{j 2 \pi m \Delta f \ell \Delta \tau}= \sum_{u=1}^U\beta_{nr,u}\mathbf{a}(\theta_{r,u}) \mathbf{a}^{\mathsf{H}}(\theta_{n,u})\, 
\mathrm{sinc}_{\tau}\left[\frac{\ell\Delta\tau - \tau_{nr,u}}{\Delta \tau}\right]
     e^{- j 2 \pi f_0 \tau_{nr,u}} 
\end{split}
\end{equation}
\hrulefill
\end{figure*} 
It is worth remarking that targets are typically non-coherent when observed by multiple APs with such angular diversity, as the complex scattering coefficient is generally shape-, frequency- and polarization-dependent. However, we herein aim at showing the trade-off between communication and imaging for a D-MIMO network, leaving the case of partial target coherence (and related processing) as a future development. In any case, the ISAC system and the methods developed in this paper are also valid for real extended targets, since the APs can be operationally clustered in subsets for which the targets can be assumed to be coherent, owing to recent literature \cite{zhi2025nearfieldintegratedimagingcommunication}. As a further remark, while targets are in far-field w.r.t. the single Tx-Rx AP pair, they are in the near-field of the D-ISAC network.

The Rx signal at the $N_{\rm rx}$ Rx APs is here assumed to be due to the scattering from the ROI\footnote{The Rx APs gather the echoes from \textit{all} targets in the environment, not just the targets in the ROI. However, as detailed in Section \ref{subsec:image_generation}, the image is generated only over the ROI, thus all the targets outside the ROI can contribute with the sidelobes of their spatial ambiguity function SAF, that are here neglected.}. Thus, at the $r$-th AP, $r\in \mathcal{N}_{\rm rx}$, we obtain the $L\times 1$ signal vector in the FT domain:
\begin{equation}\label{eq:RX_sensing_vector_FT}
\begin{split} 
\mathbf{z}_r[m,k] 
& = \sqrt{P}(1\hspace{-0.1cm}-\hspace{-0.1cm}\eta) \hspace{-0.1cm}\sum_{n \in \mathcal{N}_{\rm tx}} \hspace{-0.15cm}\mathbf{H}_{n,r}[m] \mathbf{v}^\mathrm{sen}_{n}[m] X_{n,\mathrm{sen}}[m,k] \\
& + \sqrt{P}\eta \hspace{-0.15cm}\sum_{n \in \mathcal{N}_{\rm tx}} \hspace{-0.15cm}\mathbf{H}_{n,r}[m] \sum_q \mathbf{v}^\mathrm{com}_{n,q}[m] X_{q,\mathrm{com}}[m,k] \\ & + \mathbf{n}_r[m,k]
\end{split}
\end{equation}
Here, we can represent the desired signal and the interferences similar to \eqref{eq:RX_signal_UE_FT}. The desired signal is represented by the propagation of the precoded sensing signals (at all Tx APs) through the sensing channel, while communication signals pertaining to the $Q$ data streams, are now interfering. The noise is white in time, frequency and space, $\mathbf{n}_r[m,k]\sim \mathcal{CN}(\mathbf{0}, \sigma_n^2 \mathbf{I}_L)$. It is worth noticing that, since all the APs of the D-ISAC network are operated by the same BBU, it is theoretically possible to coherently cancel the interference from the communication signal and remove its detrimental effect. However, the latter operation may be feasible if the sensing channel is perfectly known.
In practice, however, the sensing channel is unknown and composed of many paths, thus only iterative approaches can be implemented to use a first rough channel estimate in FT domain to cancel the communication signal and so on \cite{keskin2025bridginggapdataaidedsensing}. This is out of the scope of this paper, that opts for processing of the sensing signal in the DD domain, and left for a future investigation. 

In order to correctly form the single images, for each AP pair, we need to separate the contributions of different Tx APs at the $r$-th Rx AP, leveraging the orthogonality (or pseduo-orthogonality) of different sensing signals in the DD domain. Thus, the first step is to transform the FT signal \eqref{eq:RX_sensing_vector_FT}, into DD as follows:
\begin{equation}\label{eq:RX_sensing_vector_DD}
    \begin{split}
        \widetilde{\mathbf{z}}_r[\ell,p] & = 
       \sum_m \sum_k \mathbf{z}_r[m,k] e^{j 2 \pi m \Delta f \ell \Delta \tau} e^{- j 2 \pi kT p \Delta \nu} 
        \\
        & = 
        \sum_{n \in \mathcal{N}_{\rm tx}}  
        \widetilde{\mathbf{H}}_{n,r}[\ell] \otimes  \widetilde{\mathbf{s}}_{n}[\ell,p] + \widetilde{\mathbf{n}}_r[\ell,p] \\
    \end{split}
\end{equation}
where $\widetilde{\mathbf{s}}_{n,q}[\ell,p]$ is the Tx signal expressed in the DD domain, that is subject to the 2D periodic convolution with the DD channel $\widetilde{\mathbf{H}}_{n,r}[\ell]$, denoted by $\otimes$\footnote{The periodic 2D convolution between two sequences $X[n,m]$ and $Y[n,m]$, $n=0,...,N-1$, $m=0,...,M-1$ is defined as
\begin{equation}
   \{X \otimes Y\}[n,m] = \sum_{n'=0}^{N-1}\sum_{m'=0}^{M-1}  X^{*}[n',m'] Y[(n-n')\mathrm{mod}N,(m-m')\mathrm{mod}M] \nonumber
\end{equation}
where $r = a \, \mathrm{mod} \, b$ is such that $a = q b + r$, $q \in \mathbb{Z}, 0\leq r < b$. 
}. Notice that the sensing channel is function of the delay only, thus the convolution applies to the delay dimension only. The noise in the DD domain is $\widetilde{\mathbf{n}}_r[\ell,p] \in \mathcal{CN}(\mathbf{0},MK \sigma_n^2 \mathbf{I}_L)$.

The second step is to extract the sensing channel impulse response (CIR) in the DD domain by enforcing the 2D periodic correlation between the Rx signal $\widetilde{\mathbf{z}}_r[\ell,p]$ \eqref{eq:RX_sensing_vector_DD} with the $n$-th sensing signal in DD $\widetilde{X}_{n,\mathrm{sen}}[\ell,p]$.
The result is:
\begin{equation}\label{eq:est_CIR}
\begin{split}
    \widetilde{\boldsymbol{h}}_{nr}[\ell,p] &= \widetilde{\mathbf{z}}_r[\ell,p] \otimes \widetilde{X}_n[\ell,p] 
    \\
    & = \underbrace{\sqrt{P}(1\hspace{-0.1cm}-\hspace{-0.1cm}\eta)\, \widetilde{\mathbf{H}}_{n,r}[\ell] \otimes \widetilde{\mathbf{v}}^\mathrm{sen}_{n}[\ell] \otimes \widetilde{\rho}_{nn,\mathrm{sen}}[\ell,p]}_{\text{Desired signal}\;  \mathsf{S}_{nr,\mathrm{sen}}[\ell,p]} \\
    & + \underbrace{\sqrt{P}(1\hspace{-0.1cm}-\hspace{-0.1cm}\eta)\hspace{-0.1cm} \sum_{\substack {n' \in \mathcal{N}_{\rm tx} \\ n'\neq n}} \hspace{-0.1cm} \widetilde{\mathbf{H}}_{n',r}[\ell] \otimes \widetilde{\mathbf{v}}^\mathrm{sen}_{n'}[\ell] \otimes \widetilde{\rho}_{nn',\mathrm{sen}}[\ell,p]}_{\text{Sensing interference} \;  \mathsf{INT}_{nr,\mathrm{sen}}[\ell,p]} \\
    & + \underbrace{\sqrt{P}\eta \sum_{n' \in \mathcal{N}_{\rm tx}} \hspace{-0.1cm}\widetilde{\mathbf{H}}_{n',r}[\ell] \otimes \sum_q \widetilde{\mathbf{v}}^\mathrm{com}_{n',q}[\ell] \otimes \widetilde{\rho}_{n'q,\mathrm{sc}}[\ell,p]}_{\text{Communication interference} \;  \mathsf{INT}_{r,\mathrm{com}}[\ell,p]} \\
    & + \widetilde{\mathbf{d}}_r[\ell,p]
 \end{split}   
\end{equation}
where $\widetilde{\rho}_{nn,\mathrm{sen}}[\ell,p] \triangleq \widetilde{X}_{n,\mathrm{sen}}[\ell,p] \otimes \widetilde{X}_{n,\mathrm{sen}}[\ell,p]$, $\widetilde{\rho}_{nn',\mathrm{sen}}[\ell,p] \triangleq \widetilde{X}_{n,\mathrm{sen}}[\ell,p] \otimes \widetilde{X}_{n',\mathrm{sen}}[\ell,p]$ and $\widetilde{\rho}_{nq,\mathrm{sc}}[\ell,p] \triangleq \widetilde{X}_{n,\mathrm{sen}}[\ell,p] \otimes \widetilde{X}_{q,\mathrm{com}}[\ell,p]$ denote the autocorrelation of the $n$-th sensing sequence, the cross-correlation of the $n$-th and $n'$-th sensing sequence and the cross-correlation of the $n$-th sensing waveform with the $q$-th communication signal. Further, $\widetilde{\mathbf{v}}^\mathrm{sen}_{n}[\ell]$ and $\widetilde{\mathbf{v}}^\mathrm{com}_{n}[\ell]$ denote the sensing and communiction precoders in the delay domain. In \eqref{eq:est_CIR}, the first term is the desired signal for the $nr$-th AP pair, the second term is the residual interference from imperfect orthogonality among sensing waveforms, the third term is the communication interference for the sensing signal, while the last one is the white Gaussian noise after the cross-correlation with the $n$-th sensing waveform, that can be approximated as $\widetilde{\mathbf{d}}_r[\ell,p] \sim \mathcal{CN}(\mathbf{0}, \sigma_n^2 MK \mathbf{I}_L)$.

\vspace{-0.3cm}\section{Integrating Phase-Coherent Multistatic Imaging in D-MIMO Networks}\label{sec:main}

The integration of phase-coherent imaging in D-MIMO networks involves the design of the sensing waveforms and of the space-frequency precoder at each Tx AP, for both communication and sensing waveforms. The optimal selection of the Rx AP set is instead presented in Section \ref{sect:optimal_GA}.

\vspace{-0.3cm}\subsection{Sensing Waveform Design}\label{subsect:waveform_design}

Recalling Section \ref{subsec:Txsignal}, the set of sensing waveforms $\{X_{n,\mathrm{sen}}[m,k]\}$ at the Tx APs shall be designed to achieve mutual orthogonality over the ROI, and avoid the so-called \textit{MIMO noise}, i.e., the degradation of the imaging quality due to the imperfect orthogonality among different waveforms emitted at the APs. For two time-continuous generic waveforms $x_n(t)$ and $x_{n'}(t)$, they are said to satisfy the \textit{extended orthogonality condition} if: 
\begin{equation}\label{eq:extended_orthogonality}
    \int_T x_n(t) x_{n'}(t-\tau) dt = \delta_{nn'}, \,\,\, \forall \tau \in [\tau_{\min}, \tau_{\max}],
\end{equation}
i.e., their cross-correlation is zero over a finite support of delays corresponding to the minimum and maximum distance of targets in the ROI (which are only functions of the D-ISAC network geometry and ROI size and position w.r.t. network). 
For the $nn'$-th AP pair, this implies selecting the maximum and minimum delays as:
\begin{equation}
    \tau_{\min\hspace{-0.1cm}/\hspace{-0.1cm}\max, nn'} \hspace{-0.1cm} = \hspace{-0.1cm}\underset{\mathbf{x} }{\min\hspace{-0.1cm}/\hspace{-0.1cm}\max} \; \bigg\{\frac{1}{v} (\| \mathbf{x} - \mathbf{p}_n\| + \| \mathbf{p}_{n'} - \mathbf{x}\|)\bigg\}
\end{equation}
for $\mathbf{x}\in \mathrm{ROI}$. In this way, waveforms emitted by the $n$-th and $n'$-th Tx APs will be orthogonal at a given Rx AP for any scattering within the ROI. It is worth noticing that typical waveforms employed in 3GPP, such as Zadoff-Chu/Gold~\cite{andrews2025primerzadoffchusequences}, or pseudo-noise sequences used for MIMO radars, fulfill the \textit{zero-shift orthogonality}, i.e., condition \eqref{eq:extended_orthogonality} holds for $\tau=0$ only. In this latter case, waveforms are perfectly orthogonal only when the bistatic delays at a given Rx AP resulting from two Tx APs happen to be the same, resulting in image degradation in multi-target environments.

 However, selecting only the previous set of delays $[\tau_{\min, nn'},\tau_{\max, nn'}]$ ensures that only the waveforms emitted at APs $n$ and $n'$ are orthogonal, while all the others are not. In general, we need to consider the \textit{union} of all sets of delays for which we enforce the orthogonality condition, for each combination of APs:
\begin{equation}
    \mathcal{D} = \bigcup_{n,n'} [\tau_{\min, nn'},\tau_{\max, nn'}] \,.
\end{equation}
In our case, the set of delays $\mathcal{D}$ is discrete, and it applies to the samples of the \textit{periodic cross-correlation} of the waveforms, corresponding to the subset of delay samples $\widetilde{\Lambda}_\tau = \{ \ell | \ell= \lceil \frac{\tau}{\Delta \tau}\rfloor, \tau \in \mathcal{D}\}$. Herein, we design mutually orthogonal sensing waveforms in the DD domain, $\{\widetilde{X}_{n,\mathrm{sen}}[\ell,p]\}$, $n \in \mathcal{N}_{\mathrm{tx}}$, extending the methodology recently proposed in \cite{ManzoniCOSMIC} to a multistatic D-ISAC system. In order to simplify the derivation of the sensing waveforms and to reduce the complexity of the 2D cross-correlation, we design \textit{separable waveforms}, namely for which $\widetilde{X}_{n,\mathrm{sen}}[\ell,p] = \widetilde{X}^{\tau}_{n,\mathrm{sen}}[\ell] \widetilde{X}^{\nu}_{n,\mathrm{sen}}[p]$. The waveform design is carried out for the delay-related component $\widetilde{X}^{\tau}_{n,\mathrm{sen}}[\ell]$, for which we want to enforce the extended orthogonality, while the design of the Doppler-related component $\widetilde{X}^{\nu}_{n,\mathrm{sen}}[p]$ follows conventional approaches (e.g,, ZC sequences or pseudo-random sequences)\footnote{We focus on the delay-related component of sensing waveforms because all targets in this work are considered to be static, thus we need orthogonality along delay only. For moving targets, the procedure outlined here shall be extended to Doppler orthogonality as well.}.  
By stacking the samples of $\widetilde{X}^{\tau}_{n,\mathrm{sen}}[\ell]$ in a vector $\widetilde{\mathbf{x}}^\tau_{n,\mathrm{sen}} \in \mathbb{C}^{M \times 1}$, we first build the $M\times M$ periodic correlation matrix corresponding to the $n'$-th sequence, $\widetilde{\mathbf{X}}^\tau_{n',\mathrm{sen}} \in \mathbb{C}^{M \times M}$.\footnote{The periodic correlation matrix of a given sequence $\mathbf{x}\in \mathbb{C}^{M\times 1}$ is the matrix $\mathbf{X}\in \mathbb{C}^{M\times M}$ that enforces the correlation of any sequence with $\mathbf{x}$:
\begin{equation}
    [\mathbf{X}]_{i,j} = [\mathbf{x}]_{(i-j)\mathrm{mod} M}. \nonumber
\end{equation}} Then, we can enforce the extended orthogonality condition for the selected set of delay samples $\widetilde{\Lambda}_\tau$ as:
\begin{equation}\label{eq:extended_orthogonality_matrix}
    \underline{\widetilde{\mathbf{X}}}^\tau_{n',\mathrm{sen}}(\widetilde{\mathbf{x}}^\tau_{n,\mathrm{sen}})^* = \mathbf{0}
\end{equation}
where $\underline{\widetilde{\mathbf{X}}}^\tau_{n',\mathrm{sen}} \in \mathbb{C}^{|\widetilde{\Lambda}_\tau| \times M}$ denotes the cut of the correlation matrix around the $|\widetilde{\Lambda}_\tau|$ rows corresponding to $\widetilde{\Lambda}_\tau$.
Since each waveform should be orthogonal to all the others over $\widetilde{\Lambda}_\tau$, we need to apply the condition \eqref{eq:extended_orthogonality_matrix} to all pairs of waveforms. The waveform generation proceeds iteratively from $n=1$ as
\begin{equation}\label{eq:sensing_waveform_gen}
    \begin{bmatrix}
        \underline{\widetilde{\mathbf{X}}}^\tau_{n-1,\mathrm{sen}} \\
       \underline{ \widetilde{\mathbf{X}}}^\tau_{n-2,\mathrm{sen}} \\
        \vdots\\
        \underline{\widetilde{\mathbf{X}}}^\tau_{1,\mathrm{sen}}
    \end{bmatrix} \hspace{-0.1cm}(\widetilde{\mathbf{x}}^\tau_{n,\mathrm{sen}})^* \hspace{-0.1cm}= \hspace{-0.1cm}\mathbf{0} \Rightarrow \widetilde{\mathbf{x}}^\tau_{n,\mathrm{sen}} \hspace{-0.1cm} \in \mathrm{null}\hspace{-0.1cm}\left(    \begin{bmatrix}
        \underline{\widetilde{\mathbf{X}}}^\tau_{n-1,\mathrm{sen}} \\
       \underline{ \widetilde{\mathbf{X}}}^\tau_{n-2,\mathrm{sen}} \\
        \vdots\\
        \underline{\widetilde{\mathbf{X}}}^\tau_{1,\mathrm{sen}}
    \end{bmatrix}\right)
\end{equation}
where the first waveform is simply initialized as a random sequence with unit norm entries, e.g., $\widetilde{\mathbf{x}}^\tau_{1,\mathrm{sen}} \sim \mathcal{CN}(\mathbf{0}, \mathbf{I}_M)$. Thus, the $n$-th waveform is designed to lie in the null space of the stacked correlation matrices from $n=1$ to $n-1$. The size of the matrix at the $n$-th iteration is $(n-1)|\widetilde{\Lambda}_\tau| \times M$, therefore the generation is feasible if $N \leq M/|\widetilde{\Lambda}_\tau|$, i.e., when there is a sufficient null space available. As the cardinality of $\mathcal{D}$ increases with the ROI size, we have an implicit trade-off of the proposed D-ISAC system: the larger the ROI to be imaged, the less number of sensors can be used to guarantee their orthogonality. This implies that the imaging quality degrades, as shown in Section \ref{sec:results}. 
The DD sensing waveforms are then mapped in the FT domain as:
\begin{equation}
     X_{n,\mathrm{sen}}[m,k] \hspace{-0,1cm} = \hspace{-0,1cm}\sum_{\ell,p}\widetilde{X}_{n,\mathrm{sen}}[\ell,p] e^{-j 2 \pi m \Delta f \ell \Delta \tau} e^{ j 2 \pi k T p \Delta \nu }. 
\end{equation}
\textit{Remark}. The generated sensing waveforms have perfect cross-correlation properties along delay, but they are not perfectly orthogonal to the communication signals in the DD domain $\widetilde{X}_{q,\mathrm{com}}$, $q=1,...,Q$. In principle, the method in \eqref{eq:sensing_waveform_gen} can also include communication waveforms, but the procedure shall be repeated at each OFDM symbols, resulting in an overwhelming complexity. With the proposed method, the complexity is dominated by the calculation of the null-space at each iteration, for which we require $O((n-1)|\widetilde{\Lambda}_\tau| M^2)$ complex multiplications at the $n$-th iteration. This number grows quickly with $M$, but the procedure can be done \textit{offline}.
In any case, it can be demonstrated that the 2D periodic cross-correlation between the $n$-th sensing waveform and the $q$-th communication one is characterized by $\mathbb{E}[|\widetilde{\rho}_{nq,\mathrm{sc}}[0,0]|^2] = M K$, while we have $\mathbb{E}[|\widetilde{\rho}_{nn,\mathrm{sen}}[0,0]|^2] = (M K)^2$. Thus, by a sufficient number of resources $M,K$, the Rx APs can reject the communication interference.

\vspace{-0.3cm}\subsection{Space-Frequency Precoding}\label{subsect:precoding}
The design of the coherent spatial precoding vector in conventional D-MIMO communication networks is aimed at maximizing the capacity at the UEs' locations~\cite{demir2021foundations}. In our multicarrier D-ISAC system, the precoder must also be frequency-variant. However, we need to account for the illumination of the ROI as well. To this aim, we can design the spatial precoders at the $N_\mathrm{tx}$ APs to \textit{coherently focus} the signals at the $Q$ UEs and concurrently illuminate \textit{incoherently} the ROI.\footnote{For imaging, the D-MIMO network shall illuminate the \textit{entire} ROI at the same time, by leveraging proper orthogonal signals, exactly as a MIMO radar illuminates the whole environment by using antenna-orthogonal code sequences. Focusing the Tx signals in one single location within the ROI requires a 2D sweeping to form the final image, with consequent latency burden.}
The space-frequency precoding design leverages a fictitious LOS channel between the $n$-th AP and each pixel in the ROI, that we can model as
\begin{equation}
    \mathbf{h}_{n,\rm ROI} = \sum_{\mathbf{x}\in \mathrm{ROI}} \frac{\lambda_0}{4 \pi \| \mathbf{x}-\mathbf{p}_n\| }\mathbf{a}(\theta_{n}(\mathbf{x})) \,,
\end{equation}
where $\theta_{n}(\mathbf{x})$ is the AOD towards the pixel $\mathbf{x}$ of the ROI. The fictitious channel only comprises the spatial component, without any frequency-dependent behavior; any delay-induced phase shift due to scattering from targets is considered in the image formation. 

We define the composite communication channel on the $m$-th subcarrier as $\mathbf{G}[m] = [\mathbf{g}_1[m],...,\mathbf{g}_Q[m]] \in \mathbb{C}^{N_\mathrm{tx} L \times Q}$, where the $q$-th column is the channel from the Tx APs to the $q$-th UE, $\mathbf{g}_{q}[m] = [\mathbf{h}^\top_{n,q}[m], n \in \mathcal{N}_\mathrm{tx}]^\top \in \mathbb{C}^{N_\mathrm{tx} \times 1}$. Then, the spatial precoder for communication at the $N_{\mathrm{tx}}$ Tx APs is built as follows~\cite{demir2021foundations}:
\begin{equation}
    \mathbf{V}_{\rm com}[m] \hspace{-0.1cm}=\hspace{-0.1cm} \begin{dcases}
    \mathbf{G}^*[m] & \hspace{-0.1cm}\text{MR}\\
    \mathbf{G}^*[m] \hspace{-0.1cm}\left[\mathbf{G}^\mathsf{H}[m]\mathbf{G}[m] \hspace{-0.1cm}+\hspace{-0.1cm} \frac{\sigma_w^2}{P} \mathbf{I}_Q\right]^{-1} & \hspace{-0.1cm}\text{MMSE}
    \end{dcases}
\end{equation}
i.e., according to either maximum ratio (MR) or linear minimum mean squared error (MMSE) approaches, respectively. The $N_\mathrm{tx} L \times Q$ space-frequency precoder matrix is normalized according to conventional cell-free approaches so that, for $\eta=1$, the total emitted power by the D-ISAC network---per subcarrier, per UE---is $P$, thus $\mathbf{V}_{\rm com}[m] \mathbf{D}^{-\frac{1}{2}}_{\rm com}[m]$ with $\mathbf{D}_{\rm com}[m] = \mathrm{diag}(\| \mathbf{g}_{1}[m]\|^2, ..., \|\mathbf{g}_{Q}[m] \|^2)$. 

The sensing precoder is designed according to the MR principle from the composite channel from all APs to the ROI as $\mathbf{g}_{\rm ROI} = [\mathbf{h}^\top_{n,\rm ROI}, n \in \mathcal{N}_{\mathrm{tx}}]^\top \in \mathbb{C}^{N_\mathrm{tx} \times 1}$, as:
\begin{equation}
    \mathbf{V}_{\rm sen} = \mathbf{g}^*_{\rm ROI} \mathbf{1}^\top_Q.
\end{equation}
The sensing precoder is then normalized simularly to the communication one, as $\mathbf{V}_{\rm sen} \mathbf{D}^{-\frac{1}{2}}_{\rm sen}$ by $\mathbf{D}_{\rm sen} = \|\mathbf{g}_{\rm ROI}\|^2 \mathbf{I}_Q$.

In \eqref{eq:TX_signal_FT}, therefore, $\eta$ is a tunable amplitude weight that not only balances the emitted power for imaging w.r.t. communication, but also regulates the fraction of power that is used to incoherently illuminate the ROI and serve the UEs via coherent focusing. 
As in any D-MIMO system, the maximum number of UEs is bounded to be $Q \leq N_{\rm tx} L$. 

\vspace{-0.3cm}\subsection{Image Generation}\label{subsec:image_generation}
The image of the ROI is formed by space-time matched filtering (MF) on a pre-defined spatial grid $\{\mathbf{x}\}$, common to all the APs, also known as \textit{back-projection}. MF is not optimal for non-white disturbance, but it is widely adopted for its simplicity since it does not require the knowledge of the noise statistics. Moreover, it does apply to those cases where the explicit solution of the imaging problem---i.e., the estimation of the complex reflectivity of each pixel in the ROI---is difficult as the problem is ill-posed~\cite{manzoni2024wavefield}. In this work, the number of independent measures amounts to $N_{\mathrm{tx}} N_{\mathrm{rx}} L $, is typically much less than the number of pixels representing the ROI, given the achievable coherent multistatic resolution\footnote{Imaging can be cast as an inverse problem aimed at estimating the reflectivity of a map of pixels given a bunch of observations, under the \textit{linear scattering} assumption. If the number of observations is close to the number of unknowns, the inverse problem can be estimated by ZF or MMSE techniques with satisfactory results (as for tomography \cite{Wang_RTI}). In our case, the problem is too ill-posed to try a direct reconstruction from inversion, since the number of pixels is determined by the achievable multistatic resolution (centimeters) and MF is the only practical solution if not a-priori information is available.}.
Likewise, imaging with more advanced approaches, e.g., MMSE, is currently out of the scope. Since we consider static targets, images are formed from the zero-Doppler slice of the Rx signal \eqref{eq:est_CIR}, i.e., $\widetilde{\boldsymbol{h}}_{nr}[\ell] = \widetilde{\boldsymbol{h}}_{nr}[\ell,p=0]$. 
Defining $\mathbf{x}$ the generic pixel within the ROI, the single images are formed as
\begin{equation}\label{eq:image_formation_1}
\begin{split}
    I_{nr}(\mathbf{x}) & = \mathbf{a}^{\mathsf{H}}_r(\mathbf{x}) \widetilde{\boldsymbol{h}}_{nr}\left[\ell = \bigg\lceil\frac{\tau_{nr}(\mathbf{x})}{\Delta \tau}\bigg\rfloor\right] e^{+ j 2 \pi f_0 \tau_{nr}(\mathbf{x})} \\ & = \sum_{u=1}^U \xi_{nr,u} \chi_{nr}(\mathbf{x}-\mathbf{x}_u) + D_{nr}(\mathbf{x}),
\end{split}
\end{equation}
where $\mathbf{a}_r(\mathbf{x})$ is the Rx steering vector of the $r$-th Rx AP referred to the pixel $\mathbf{x}$ and $\tau_{nr}(\mathbf{x})$ is the two-way delay from the center of the $n$-th Tx AP to the center of the $r$-th Rx AP, computed by substituting $\mathbf{x}$ to $\mathbf{x}_u$ in \eqref{eq:delay}. The disturbance $D_{nr}(\mathbf{x})$ is the mapping of the noise and interference in \eqref{eq:est_CIR} in the spatial domain. The image formation consists of the compensation of propagation phase terms induced by the propagation from the center of the $n$-th Tx AP to each antenna element of the $r$-th Rx AP, and the evaluation of the CIR at the delay sample corresponding to the two-way delay from pixel $\mathbf{x}$. The result of the formation of single images is the collection of scaled and shifted replicas of the \textit{spatial ambiguity function (SAF)} of the $nr$-th AP pair $\chi_{nr}(\mathbf{x})$, that is the image of a point target as seen by the $nr$-th AP pair. The noise in the spatial domain is $D_{nr}(\mathbf{x})$. The SAF is the spatial dual of the well-known DD ambiguity functions of radars, and its formal definition is left to the available literature for brevity~\cite{manzoni2024wavefield,tagliaferri2024cooperative}.

The coherent image is then obtained as:
\begin{equation}\label{eq:image_tot}
    I(\mathbf{x}) = \sum_{\substack{n \in \mathcal{N}_{\rm tx} \\ r \in \mathcal{N}_{\rm rx}} } I_{nr}(\mathbf{x}) = \sum_{u=1}^U \xi_u \chi(\mathbf{x}-\mathbf{x}_u)  + D(\mathbf{x})
\end{equation}
again a collection of \textit{coherent SAFs} $\chi(\mathbf{x})$, i.e., the image of a point target achieved by the entire D-ISAC (only the selected Tx-Rx pairs). The coherent SAF is characterized by a much narrower main lobe (thus a higher theoretical resolution) but also a higher level of sidelobes, due to the spatial gaps among APs. Since any image can be represented as the coherent combination of multiple replicas of the SAF, the imaging capability of the proposed D-ISAC system are fully characterized by the analysis of the properties of the SAF.


\vspace{-0.3cm}\section{Performance Metrics}\label{sec:metric}

\subsection{Communication Metric}
The main performance indicator for communication is the SE at the $q$-th UE, averaged over the DL burst. The expression of the SE can be approximated (for white Gaussian disturbance) as: 
\begin{equation}\label{eq:SE}
    \mathsf{SE}_q \simeq \frac{1}{MK} \sum_k \sum_{m}  \log_2 \left(1 +  \mathsf{SINR}^{\rm com}_{q}[m,k]\right)
\end{equation}
where the SINR at $q$-th UE over the $mk$-th FT resource bin is shown in \eqref{eq:SINR_com}. The SINR is obtained from \eqref{eq:RX_signal_UE_FT} as the ratio of the power of the desired signal $\mathsf{S}_{q,\rm com}[m,k]$ and the overall interference plus noise, composed of $\mathsf{INT}_{q,\mathrm{com}}[m,k]$, $\mathsf{INT}_{q,\mathrm{sen}}[m,k]$ and $W_q[m,k]$. The expression of the SINR, assuming LOS channels between APs and UEs, is shown in \eqref{eq:SINR_com}. Herein, $F^{\rm com}_{n,qq'}[m] = \mathbf{a}^\top(\theta_{nq'})\mathbf{v}^{\rm com}_{n,q}[m]$ is the frequency-dependent spatial precoding gain of the communication signal emitted by the $n$-th Tx AP to the $q'$-th UE, for $q'\neq q$ (not the intended UE), while $F^{\rm sen}_{n,q}[m] = \mathbf{a}^\top(\theta_{nq'})\mathbf{v}^{\rm sen}_{n,q}[m]$ denotes the frequency-dependent spatial precoding gain of the sensing signal emitted by the $n$-th Tx AP at the $q$-th UE location. Factor $\overline{\alpha}_q = \| \mathbf{g}_q\|$ accounts for the normalization of the spatial precoders by the norm of the composite communication channel at the $N_{\rm tx}$ APs. The upper bounds is represented by the D-MIMO system ($\eta=1$) for perfect MUI cancelation (perfect CSI and MMSE precoding). 
\begin{figure*}
    \begin{equation}\label{eq:SINR_com}
    \begin{split}
        \mathsf{SINR}^{\rm com}_{q}[m,k]\simeq \frac{\displaystyle  P \eta^2 \sum_{n \in \mathcal{N}_{\rm tx}} \bigg\lvert \frac{\alpha_{nq}}{\overline{\alpha}_q} \bigg\rvert^2 }{\displaystyle P \eta^2 \sum_{q'\neq q} \sum_{n \in \mathcal{N}_{\rm tx}} \bigg \lvert \frac{\alpha_{nq'} }{\overline{\alpha}_q'} F^{\rm com}_{n,qq'}[m] \bigg\rvert^2 + P (1-\eta)^2\sum_{n \in \mathcal{N}_{\rm tx}} \bigg\lvert \frac{\alpha_{nq}}{\overline{\alpha}_q} F^{\rm sen}_{n,q}[m]\bigg\rvert^2  + \sigma_w^2}  \leq \underbrace{\frac{\displaystyle P \sum_{n =1}^N \bigg\lvert \frac{\alpha_{nq}}{\overline{\alpha}_q} \bigg\rvert^2 }{\displaystyle \sigma^2_{w} }}_{\mathsf{SNR}^{\rm D-MIMO}_{q}}
    \end{split}
\end{equation}
\hrulefill
\end{figure*}



\vspace{-0.5cm}\subsection{Sensing Metrics}


\subsubsection{SINR} The analysis of the sensing SINR for the single $nr$-th Tx-Rx AP pair allows identifying the role of the various disturbance terms. We can evaluate an approximation of the sensing SINR considering one target in the ROI, located in $\overline{\mathbf{x}}$. The expression is reported in \eqref{eq:sensing_SINR_extended}.
\begin{figure*}[!t]
\begin{equation}\label{eq:sensing_SINR_extended}
\begin{split}
\mathsf{SINR}^{\rm sen}_{nr} 
        & \simeq 
        \frac{\displaystyle  P (1-\eta)^2 MK |\beta_{nr}|^2 }{\displaystyle P (1-\eta)^2  \sum_{\substack{n' \in \mathcal{N}_{\rm tx}\\ n'\neq n}} |\beta_{n'r}|^2  + P \eta^2  \sum_{\substack{n' \in \mathcal{N}_{\rm tx}\\ n'\neq n}} |\beta_{n'r}|^2 \sum_{q=1}^Q  \lvert \widetilde{F}_{n',q} \rvert^2 + \sigma_n^2/L} 
        \leq \underbrace{\frac{ \displaystyle P MK |\beta_{nr}|^2   L}{\displaystyle  \sigma_n^2 }}_{\mathsf{SNR}^{\rm D-RN}_{nr} }
\end{split}
\end{equation}
\hrulefill
\end{figure*}
Herein, the numerator combines the sensing channel gain in DD in its \textit{peak} $|\beta_{nr}|^2$, the fraction of power allocated for sensing $(1-\eta)^2$, the matched filter gain in DD, $MK$. At the denominator, the first term is the residual interfering contribution of the other emitted sensing waveforms, in case of imperfect orthogonality (this term is zero for the waveforms generated according to Section \ref{subsect:waveform_design}). The second term at the denominator is the power of the interfering contribution from communication signals over the $Q$ data streams, where $\widetilde{F}_{n,q} = \mathbf{a}^{\mathsf{T}}(\theta_{n}(\overline{\mathbf{x}})) \widetilde{\mathbf{v}}^{\rm com}_{n,q}[\ell]$ is the beamforming gain for the communication signal emitted by the $n$-th Tx AP intended to the $q$-th UE, evaluated in the direction of the target. Notice that the frequency-dependend precoder employed for communication is mapped into a delay-depended precoder, that is evaluated in the very same peak position of the desired signal power is present to obtain $\widetilde{F}_{n,q}$. The last term at the denominator is the thermal noise power after DD matched filtering and spatial combining at the Rx side, whose power is abated by $L$. The sensing SINR is upper bounded (for equal emitted power at the Tx APs $P$) by the D-RN case, where APs are capable of full-duplex operation and emit only the sensing waveforms (thus $\eta \rightarrow 0$). The SNR upper bound is then achieved only for perfect waveform orthogonality among the APs. 

\subsubsection{Shannon entropy of the SAF} The quality of the coherent SAF can be measured in terms of its Shannon entropy, measuring the average information content of the SAF itself. A lower entropy means that the energy of the SAF is concentrated in the main lobe and less in the sidelobes. Shannon entropy is one of the main metrics used to evaluate the focusing of arbitrary radio images~\cite{e24040509} (along with image contrast~\cite{532282}), and can apply to \textit{any} targets' distribution in space, resulting in an effective figure of metrit for image quality\footnote{Other metrics such as peak-to-sidelobe ratio (PSLR) and integrated sidelobe level (ISLR) can be used as well, but Shannon entropy can also be used as a proxy for image quality with an arbitrary number of targets.}. Therefore, 
by defining the normalized intensity image as $ \underline{I}(\mathbf{x}) = |I(\mathbf{x})|^2 / \left(\sum_{\mathbf{x}\in \mathrm{ROI}} |I(\mathbf{x})|^2 \right)$, the Shannon entropy is~\cite{e24040509}:
\begin{equation}
    \mathsf{E}(I) = - \sum_{\mathbf{x}\in \mathrm{ROI}} \underline{I}(\mathbf{x}) \log_2 \underline{I}(\mathbf{x}).
\end{equation}
To gain some insights, consider a SAF represented by a 2D sinc function $I(\mathbf{x}) = I_0 \, \mathrm{sinc}[x/\rho_x] \mathrm{sinc}[y/\rho_y]$, with spatial resolution $\rho_x$ and $\rho_y$ along $x$ and $y$, respectively. It can be demonstrated that the entropy can be approximated as $\mathsf{E}(I) \approx \log_2(\kappa \rho_x) + \log_2(\kappa \rho_y)$, with $\kappa\approx1.36$, where we used the continuous definition of the entropy (valid for very small pixels) and approximate the integral by the effective size of the main lobe. The more $\rho_x$ and $\rho_y$ reduce, the smaller is the Shannon entropy, meaning that the SAF energy is concentrated in a smaller area (more focused). In the limit for which $\rho_x$ and $\rho_y$ approach the pixel size, the Shannon entropy goes to zero. For a geenric image composed of $U$ well-separated targets, we have $\mathsf{E}(I) \approx U \left(\log_2(\kappa \rho_x) + \log_2(\kappa \rho_y)\right)$, while for dense scenarios, the grow of the enropy is sub-linear.
We use the entropy to both evaluate the ISAC trade-offs in the proposed system and enforce the optimization problem aimed at selecting the optimal Rx APs. 

\vspace{-0.3cm}\section{Optimized Selection of Rx APs}\label{sect:optimal_GA} 

The selection of the Rx set of APs proposed herein is aimed at maximizing the \textit{effective resolution} of the D-ISAC network, i.e., achieving a sharp mainlobe for the SAF and low sidelobes. The unconstrained problem (i.e., a pure sensing system with half-duplex nodes) would select half of the APs as Tx and half of Rx, to maximize the number of multistatic pairs---$N_{\rm tx}N_{\rm rx}/2$, maximized when $N_{\rm rx}=N_{\rm tx}=N/2$. This yields the best possible \textit{spectral coverage}, i.e., the largest possible set of spatial frequencies composing the final coherent SAF\footnote{The spectral coverage is the set of spatial frequencies proper of the target than can be observed by the multistatic D-ISAC configuration, and it is primarily function of the number of multistatic pairs and the ISAc nodes' position in space~\cite{manzoni2024wavefield}. The SAF is the 2D inverse Fourier transform of the spectral coverage, where each tile is properly weighted by path-loss, precoding gain etc. To a minor extent, the employed bandwidth and number of antennas play a role.}. However, increasing the number of Rx APs decreases the SE, as shown in Section \ref{sec:results}, thus the selection problem herein proposed considers an active constraint on the maximum number of allowed Rx APs. 

The selection of $\mathcal{N}_{\rm rx}$ Rx APs is carried out by minimizing the entropy of the coherent image for a single target located in the center of the ROI $\overline{\mathbf{x}}$ (the SAF). Then, the specific Rx AP set change according to the relative position of the ROI w.r.t. ISAC D-MIMO network. Given the $N$ APs, we can define two Boolean vectors $\boldsymbol{b}_{\rm tx}\in\mathbb{B}^{N \times 1}$, $\boldsymbol{b}_{\rm rx}\in\mathbb{B}^{N \times 1}$, with the following meaning: if $[\boldsymbol{b}_{\rm tx}]_{n} = 1$ means that the $n$-th AP is Tx, and zero otherwise. Similarly, $[\boldsymbol{b}_{\rm rx}]_{r} = 1$ means that the $r$-th AP is operating as Rx, and zero otherwise. Before formalizing the optimization problem, we rewrite the expression of the coherent single-target image (the SAF) as a function of $\boldsymbol{b}_{\rm tx}$ and $\boldsymbol{b}_{\rm rx}$:
\begin{equation}
    I(\mathbf{x};\boldsymbol{b}_{\rm tx},\boldsymbol{b}_{\rm rx}) =  \sum_{n\in \mathcal{N}} \sum_{r\in \mathcal{N}} [\boldsymbol{b}_{\rm tx} ]_n [\boldsymbol{b}_{\rm rx} ]_r I_{nr}(\mathbf{x}).
\end{equation}
where $I_{nr}(\mathbf{x})$ is obtained as \eqref{eq:image_tot} for a single target in $\overline{\mathbf{x}}$.
The Rx AP allocation problem can be formally posed as follows:
\begin{subequations}\label{eq:optProb4}
\begin{alignat}{2} 
&\underset{\boldsymbol{b}_{\rm tx},\boldsymbol{b}_{\rm rx}}{\mathrm{min}}   &\quad&   
- \sum_{\mathbf{x}} \underline{I}(\mathbf{x};\boldsymbol{b}_{\rm tx},\boldsymbol{b}_{\rm rx}) \log_2 \left(\underline{I}(\mathbf{x};\boldsymbol{b}_{\rm tx},\boldsymbol{b}_{\rm rx})\right)\label{eq:optProb3_obj}\\ 
&   \mathrm{s.\,t.}&      & [\boldsymbol{b}_{\rm tx} ]_n + [\boldsymbol{b}_{\rm rx} ]_n \leq 1, \,\,\,\, n=1,...,N
\label{eq:optProb3_c1}\\
&  &      &\sum_{n=1}^N [\boldsymbol{b}_{\rm rx} ]_n \leq N^{\rm max}_{\rm rx}
\label{eq:optProb3_c4}\\
&  &      &\sum_{n=1}^N [\boldsymbol{b}_{\rm tx} ]_n + [\boldsymbol{b}_{\rm rx} ]_n =N
\label{eq:optProb3_c5}
\end{alignat}
\end{subequations}
where the objective \eqref{eq:optProb3_obj} is the entropy minimization, the first constraint \eqref{eq:optProb3_c1} enforces the half-duplex setting, in which an AP cannot be both Tx and Rx at the same time, constraint \eqref{eq:optProb3_c4} sets a maximum number of Rx APs, to limit the SE degradation in practice, while \eqref{eq:optProb3_c5} imposes the usage of all the APs, either Tx or Rx (no idle). The solution of problem \eqref{eq:optProb3_obj} primarily depends on the geometry of the D-ISAC network and the employed bandwidth, but also---on a minor extent---on the relative UEs-ROI position. The problem is non-convex w.r.t. $\boldsymbol{b}_{\rm tx}$ and $\boldsymbol{b}_{\rm rx}$ and not easy to solve with standard approaches. Since the allocation procedure can be carried out offline, we consider solving \eqref{eq:optProb3_obj} with a genetic algorithm (GA).

\begin{figure}[!t]
    \centering
    \includegraphics[width=0.9\linewidth]{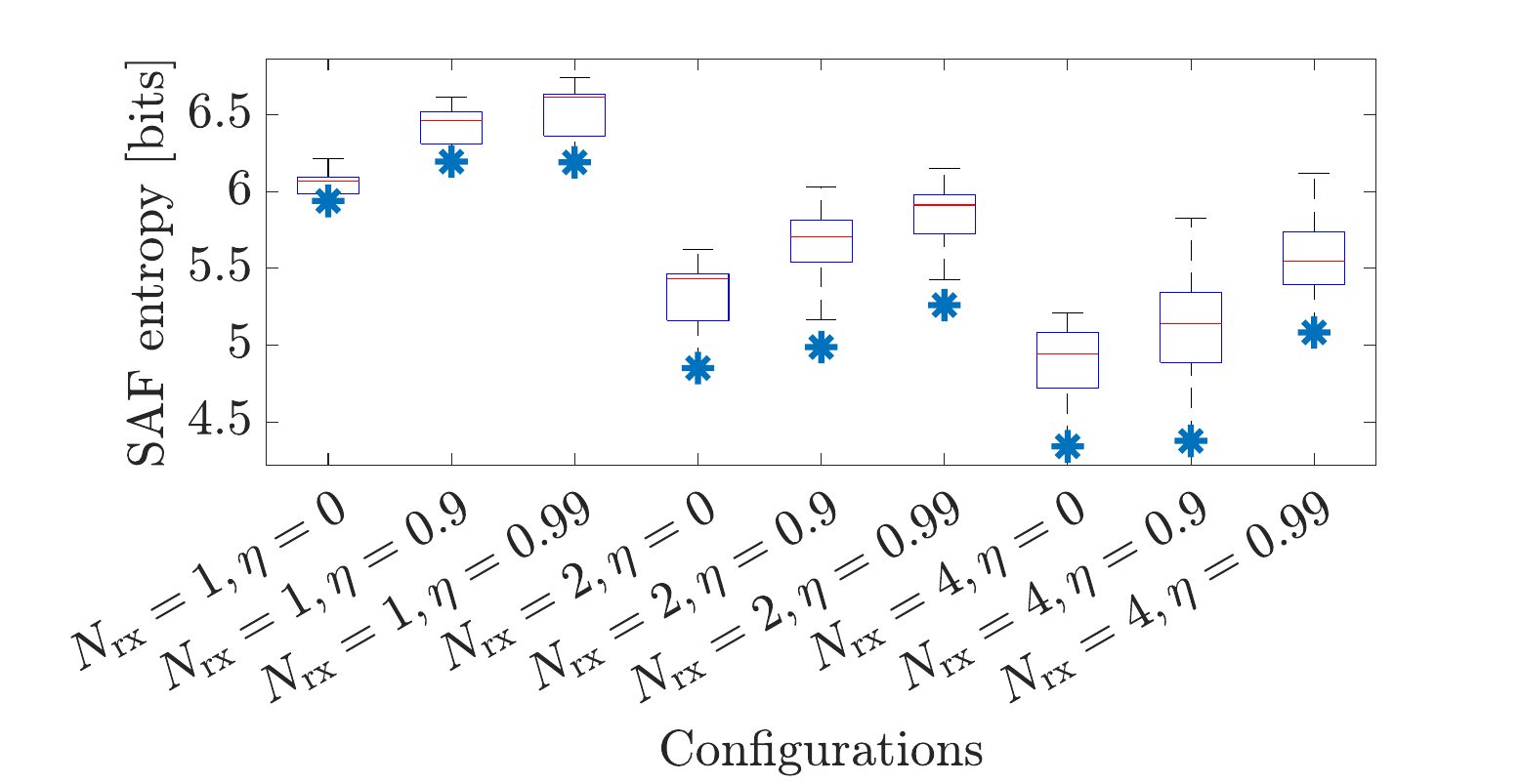}
    \caption{Distribution Shannon entropy of the SAF (i.e., for random Rx AP allocation) varying the trade-off factor $\eta$ and the number of Rx APs $N_{\rm rx}$. Optimal---minimum---values are achieved solving problem \eqref{eq:optProb3_obj} (blue stars).}\label{fig:GA}
\end{figure}

\vspace{-0.3cm}\section{Numerical Results}\label{sec:results}

This section shows and discusses some results aiming at demonstrating the feasibility of the proposed imaging system in downlink D-MIMO networks. 

\vspace{-0.3cm}\subsection{Scenario and Benchmarks}\label{subsect:scenario}

We consider a D-ISAC topology with multiple configurations. We consider $N\in\{4,9,16\}$ APs regularly deployed over a $20 \times 20$ m$^2$ area, each equipped with $L\in\{9,4,2\}$ antennas, respectively. The D-ISAC network operates at carrier frequency $f_0=10$ GHz with bandwidth $B\in\{100,1000\}$ MHz and it serves $Q=2$ UEs while imaging a $5 \times 5$ m$^2$ ROI where one target is present in the center. The single target case, i.e., the generation of the SAF, eases the evaluation of the image quality and its interpretation. The multi-target image follows from the SAF with \eqref{eq:image_formation_1}.
The Tx APs employs a FT resource grid of variable size, within the sets $M \in \{128,512,2048,4096\}$ and $K\in\{1,4\}$ and the emitted power per-subcarrier is $P = -15$ dBm@$M=128$ (the power scales with $M$ to yield a constant power spectral density). Unless otherwise specified, changing $M$ means changing the subcarrier spacing $\Delta f$, as the bandwidth $B$ is fixed. The $M,K$ configuration depends on the specific set of results, as discussed. The number of Rx APs changes in the set $N_{\rm rx} \in \{1,2,4,8\}$, depending on the configuration, and they are selected with the strategy in Section \ref{sect:optimal_GA}. Similarly, the results are averaged over random UE and target positions. The signal at UEs and at Rx APs is corrupted by thermal noise with power spectral density of $-173$ dBm/Hz. We compare the performance of our proposed D-ISAC system with the following benchmarks:
\begin{enumerate}
    \item \textbf{D-MIMO}. The APs are all used as Tx to serve the UEs, emitting only the communication signal. Therefore, there is no sensing signal and \mbox{$\eta=1$}. This represents the performance upper bound for communication~\cite{demir2021foundations}.
    \item \textbf{D-RN}. The APs operate in \textit{full-duplex} to configure a full multistatic D-RN to optimally image the ROI. Therefore, there is no communication signal and \mbox{$\eta=0$}. This represents the performance upper bound for sensing~\cite{tagliaferri2024cooperative}.
\end{enumerate}

 \begin{figure}
    \centering
    \subfloat[][SE]{\includegraphics[width=0.91\linewidth]{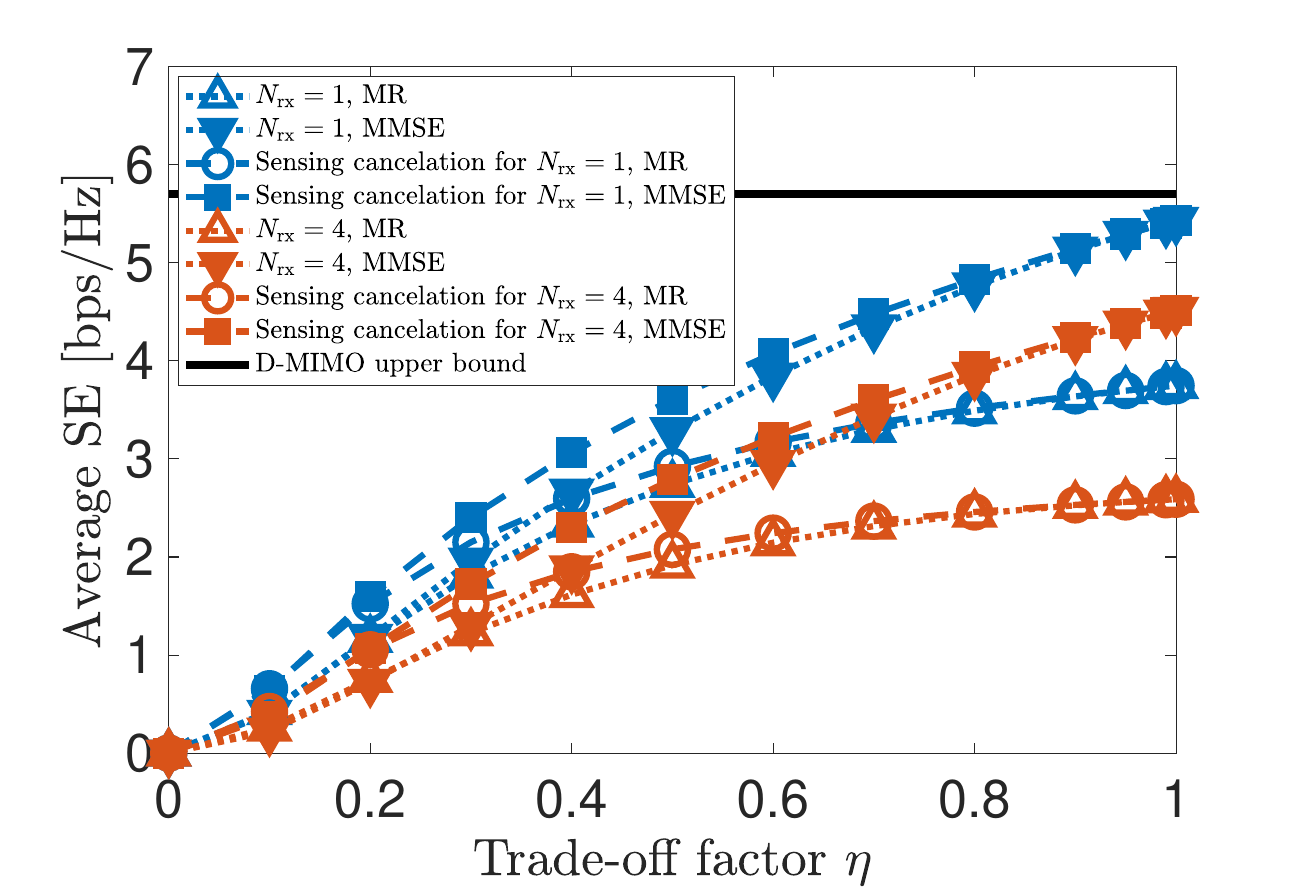}\label{subfig:SE_VS_eta_Q=2_B=100MHz}}\vspace{-0.4cm}
    \\
    \subfloat[][SINR]{\includegraphics[width=.91\linewidth]{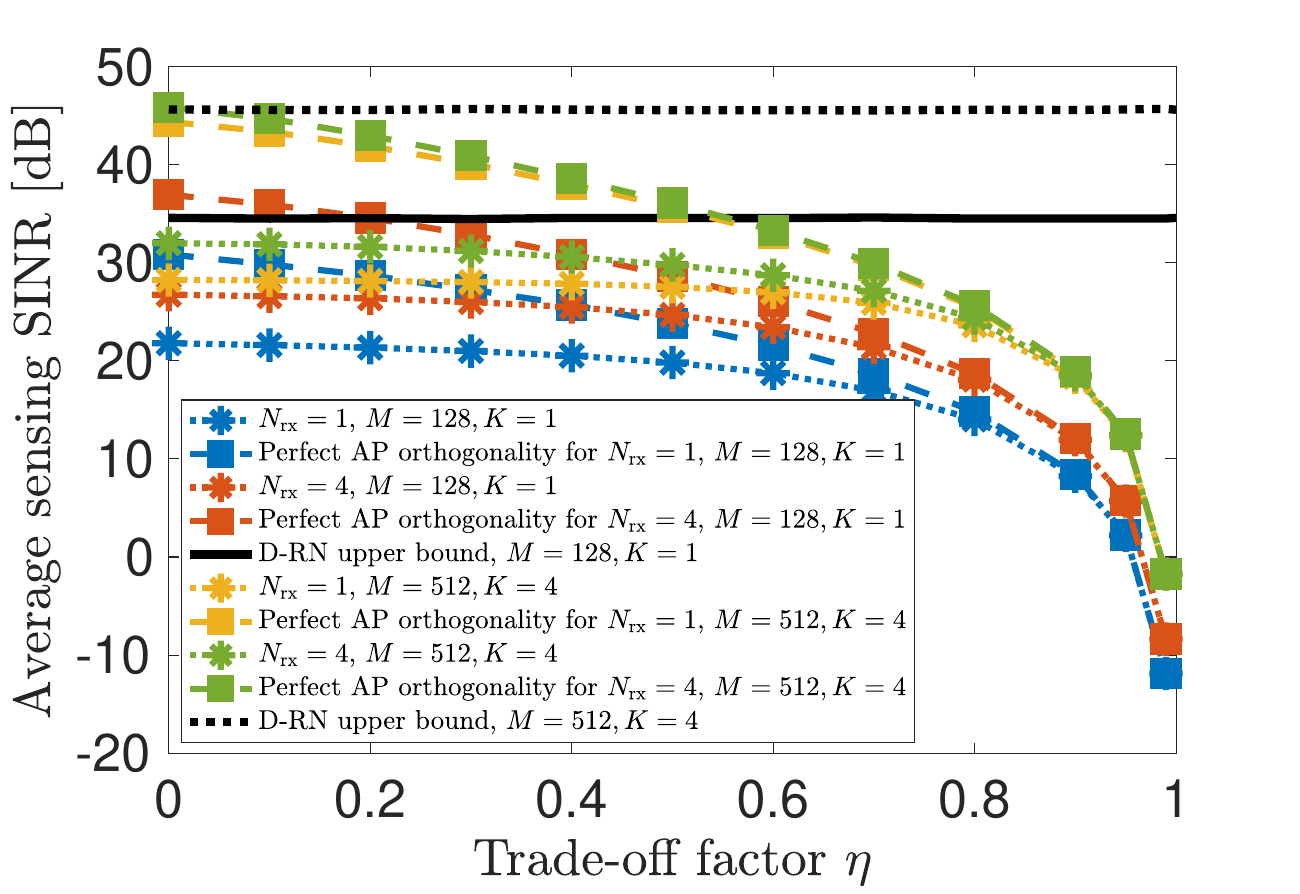}\label{subfig:SINR_VS_eta_Q=2_B=100MHz}}\vspace{-0.4cm}
        \\
    \subfloat[][SAF entropy]{\includegraphics[width=.91\linewidth]{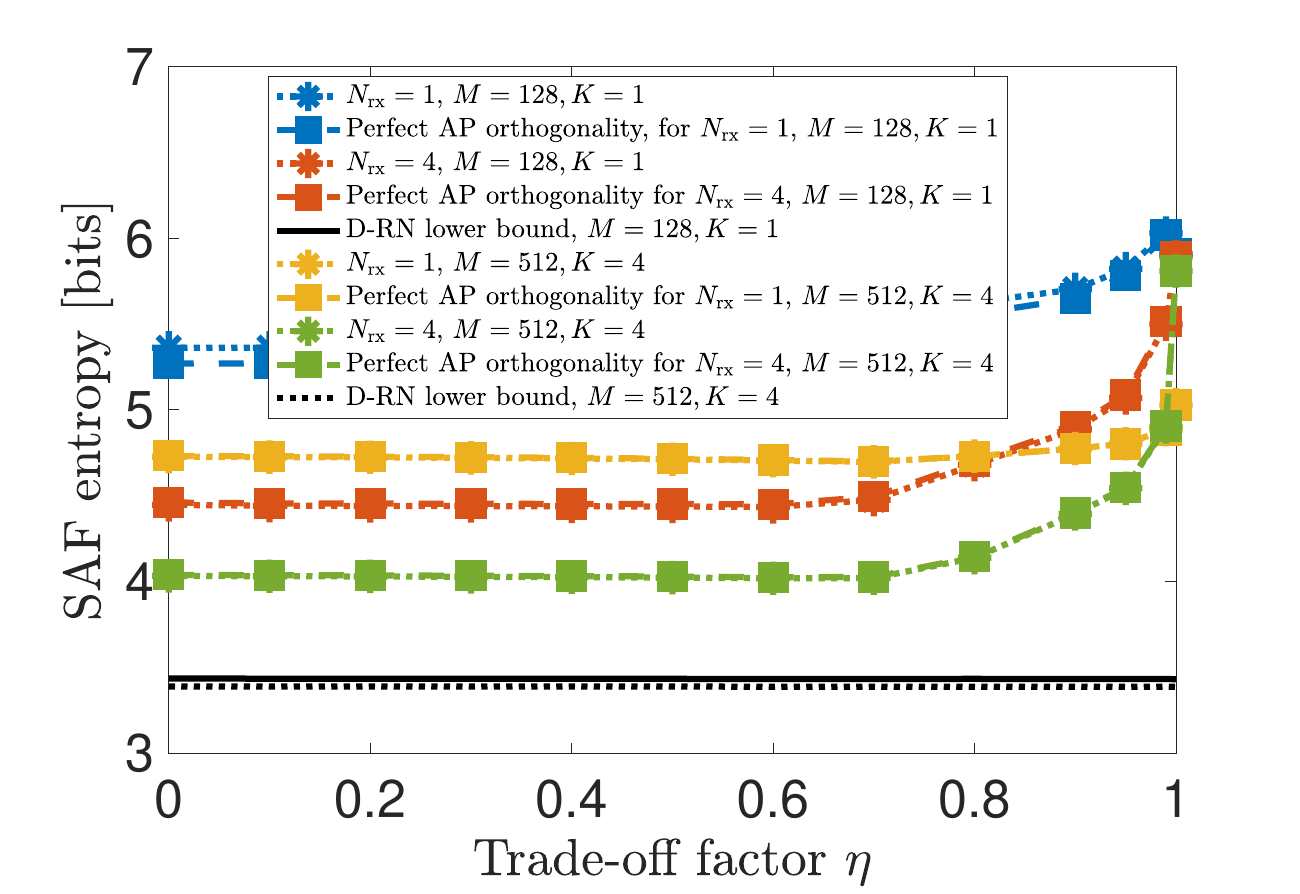}\label{subfig:ENTROPY_VS_eta_Q=2_B=100MHz}} 
    \caption{D-ISAC performance for $N=9$ APs, $L=4$ antennas, $B=100$ MHz: (a) SE, (b) SINR, (c) Entropy of the SAF. }
    \label{fig:KPI_vs_eta}
\end{figure}

\vspace{-0.3cm}\subsection{Optimal vs. Random Rx AP allocation}\label{subsect:performance_optimal_vs_random_alloc}

 An example of application of GA to problem \eqref{eq:optProb3_obj} is shown in Fig. \ref{fig:GA}, obtained for $N=9$ APs, $L=4$ antennas each, and $M=512,K=4$. Interestingly, the achievable SAF entropy gain w.r.t. random Rx AP selection increases with $N_{\rm rx}$. Although not shown for brevity, the minimum entropy is achieved for $N_{\rm rx} = N/2$, i.e., when half of the APs are Rx. This is the configuration that maximizes the number of multistatic AP pairs $(N-N_{\rm rx})N_{\rm rx}/2$, thus the spectral coverage and, in turn, imaging quality, but the impact on the D-MIMO system when using half of APs as Rx is significant in terms of SE. Therefore, we are interested in those configurations for which the number of Rx APs is less than $25\%$ of the total. For $N_{\rm rx} = 1$ and $\eta=0$ (no communication signal, sensing-centric system), the gain is minimal, i.e., the performance is limited by the D-ISAC topology. However, increasing $\eta$ increases the absolute entropy (mainly due to communication interference) but an optimal Tx-Rx AP allocation becomes more beneficial. For $N_{\rm rx} = 4$, the optimized vs. random selection gap is maximal. Noticeably, the achievable entropy for $\eta=0.9$ is very similar to the entropy for $\eta=0$ (a sensing-centric system). 

\begin{figure*}
\centering
\subfloat[][D-RN]{\includegraphics[width=0.25\linewidth]{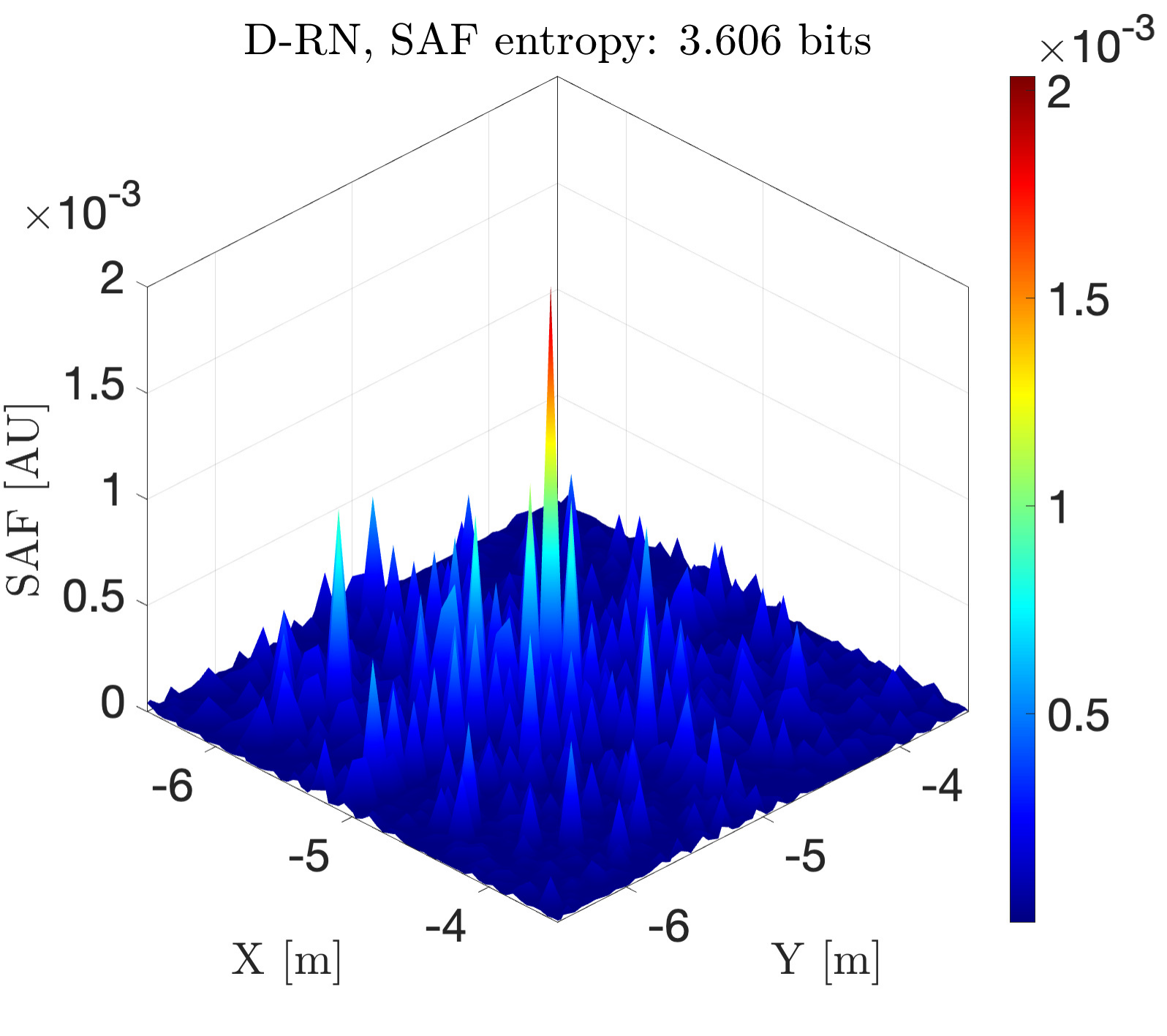}\label{subfig:SE_vs_eta}}
\subfloat[][$\eta=0$ ]{\includegraphics[width=0.25\linewidth]{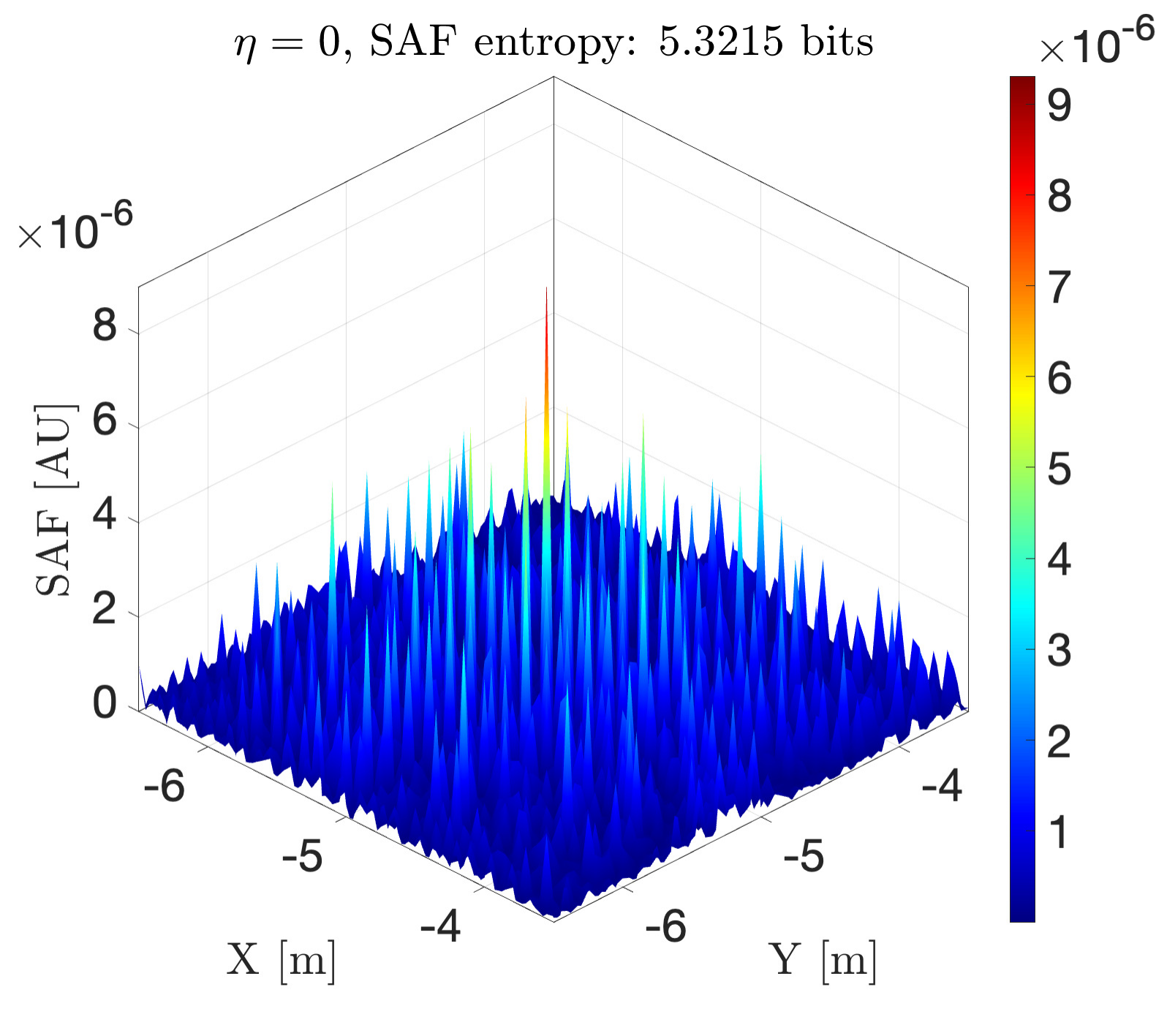}\label{subfig:SINR_vs_eta}}
\subfloat[][$\eta=0.95$ ]{\includegraphics[width=0.25\linewidth]{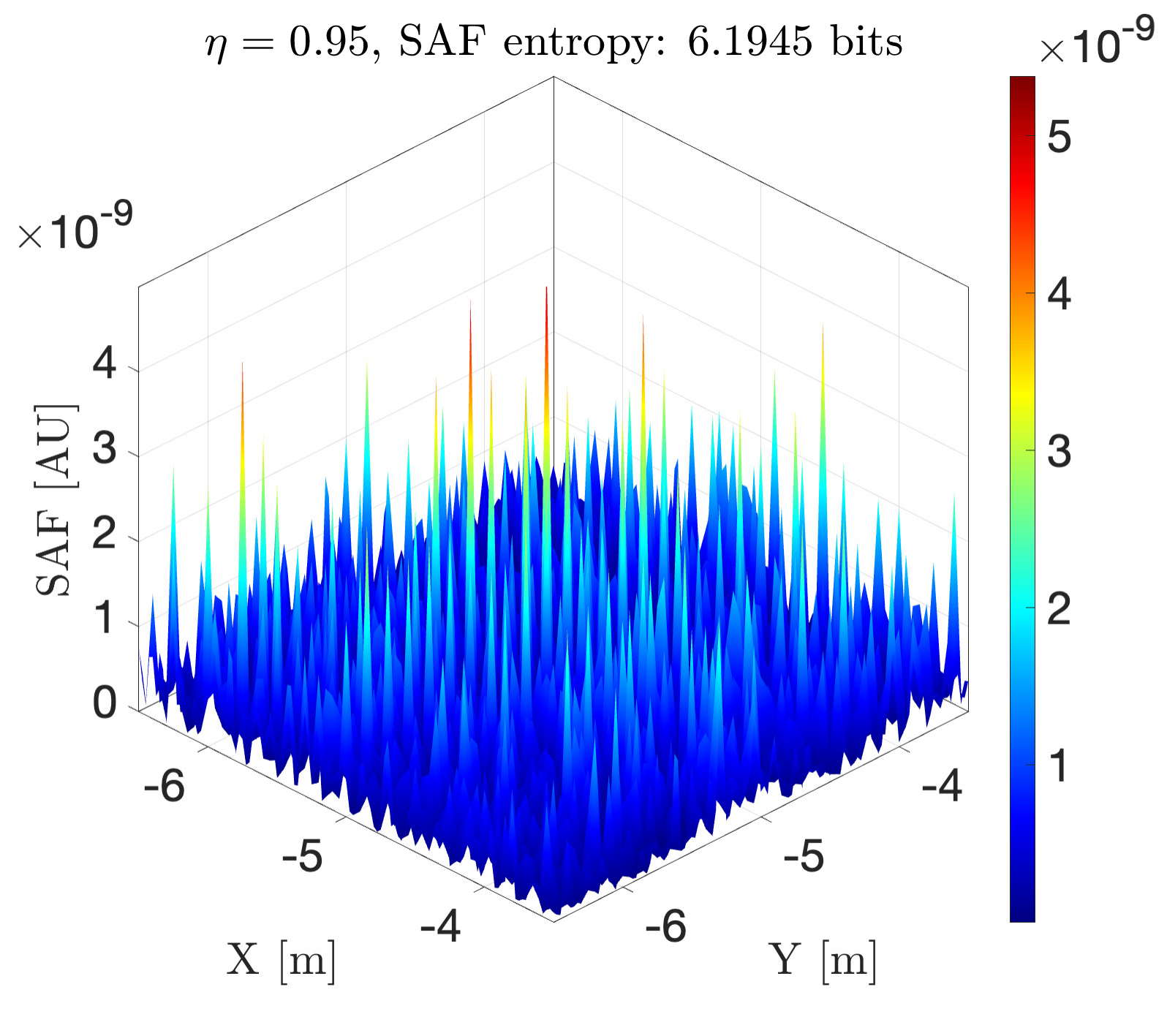}\label{subfig:entropy_vs_eta}}
\subfloat[][$\eta=0.99$]{\includegraphics[width=0.25\linewidth]{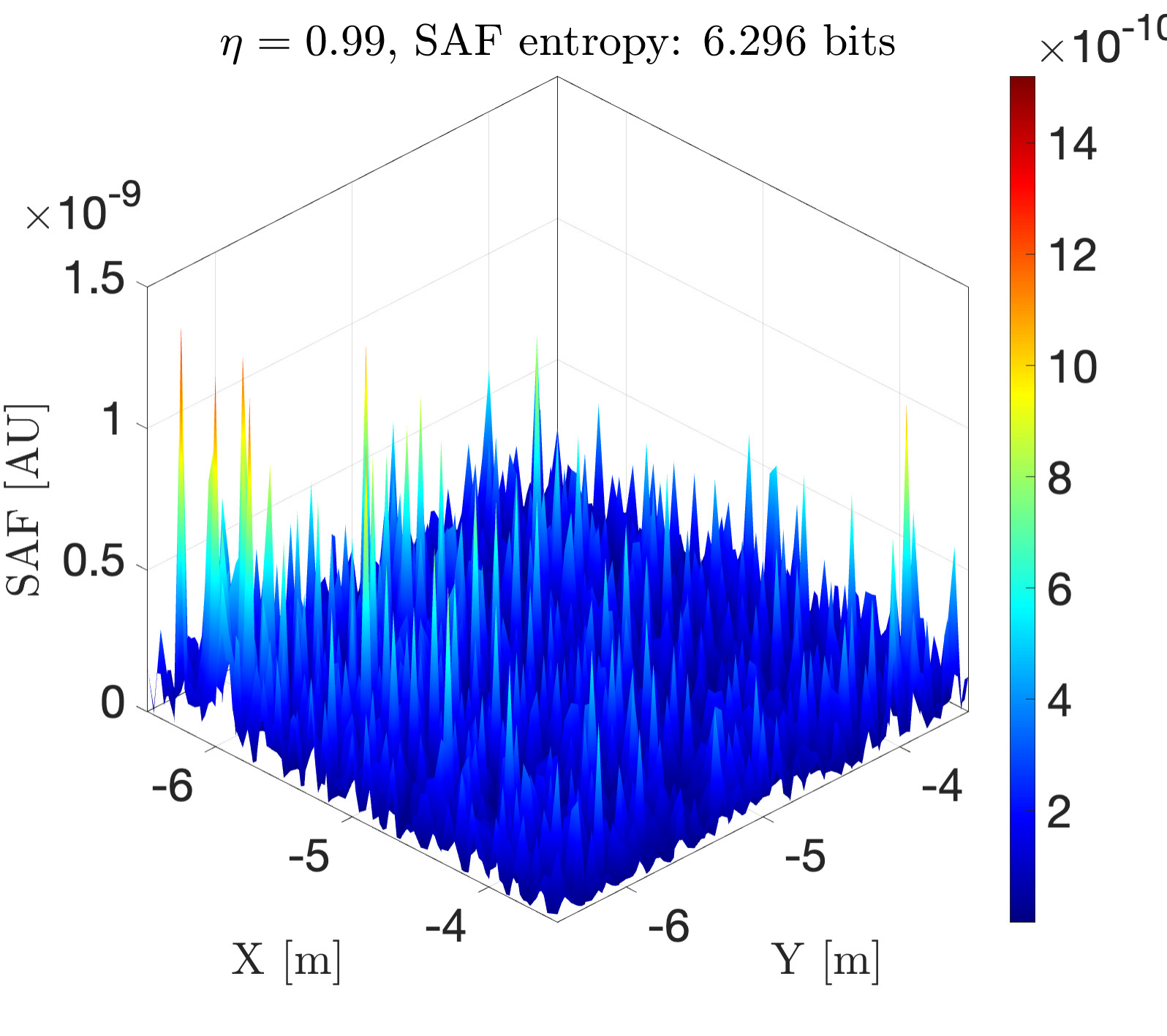}\label{}}\vspace{-0.3cm}\\


\subfloat[][D-RN]{\includegraphics[width=0.25\linewidth]{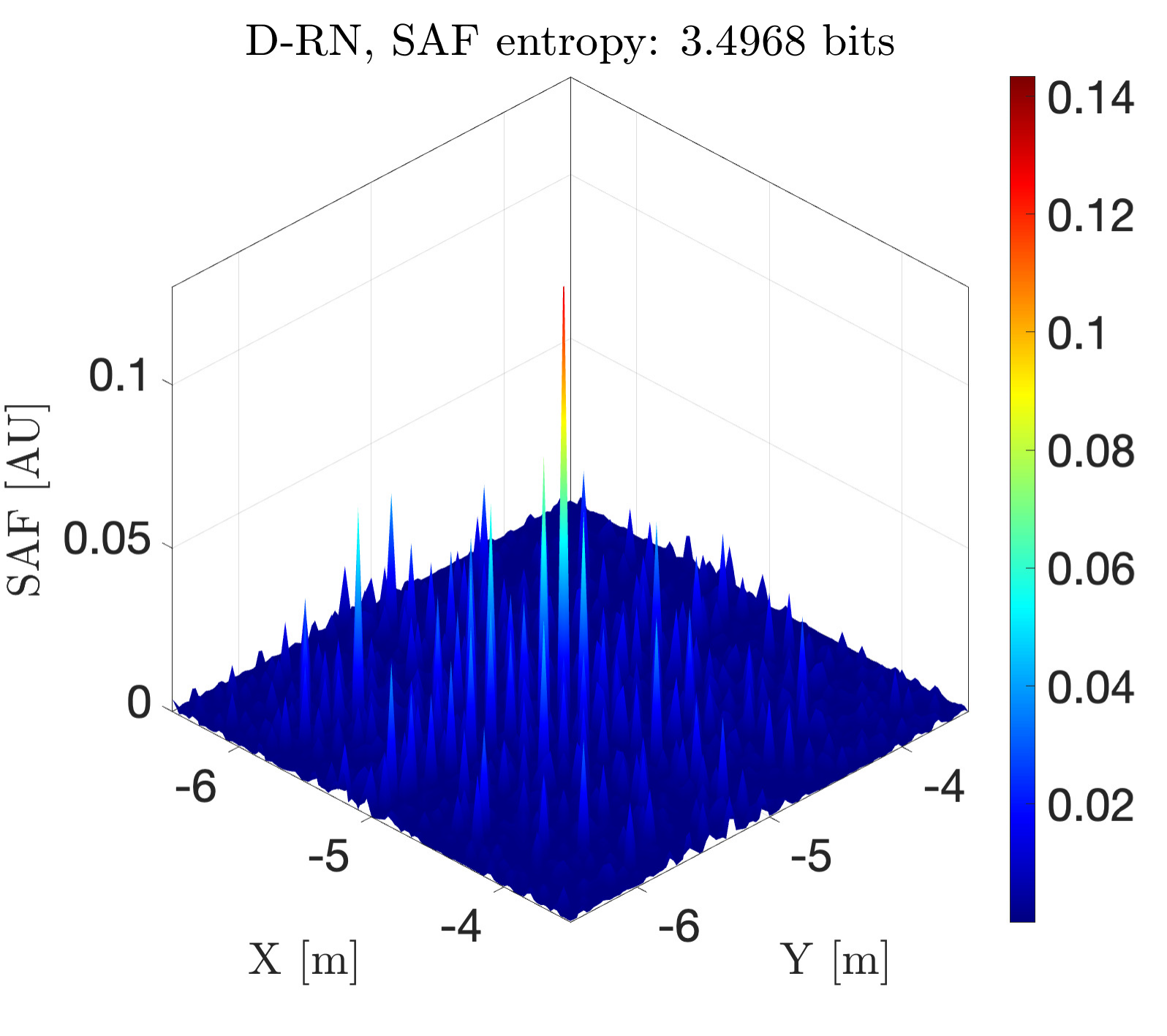}\label{subfig:SE_vs_eta}}
\subfloat[][$\eta=0$ ]{\includegraphics[width=0.25\linewidth]{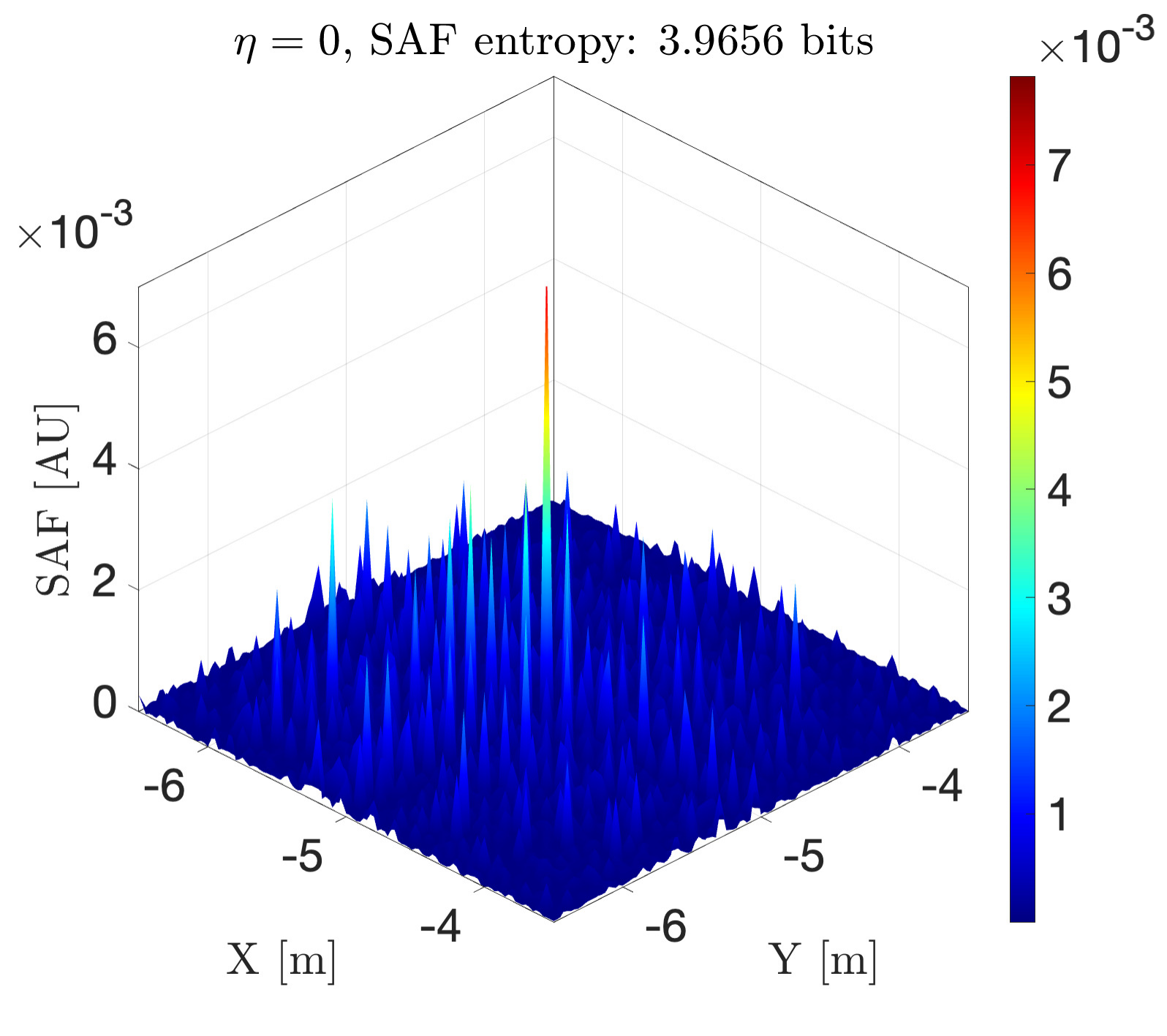}\label{subfig:SINR_vs_eta}}
\subfloat[][$\eta=0.95$ ]{\includegraphics[width=0.25\linewidth]{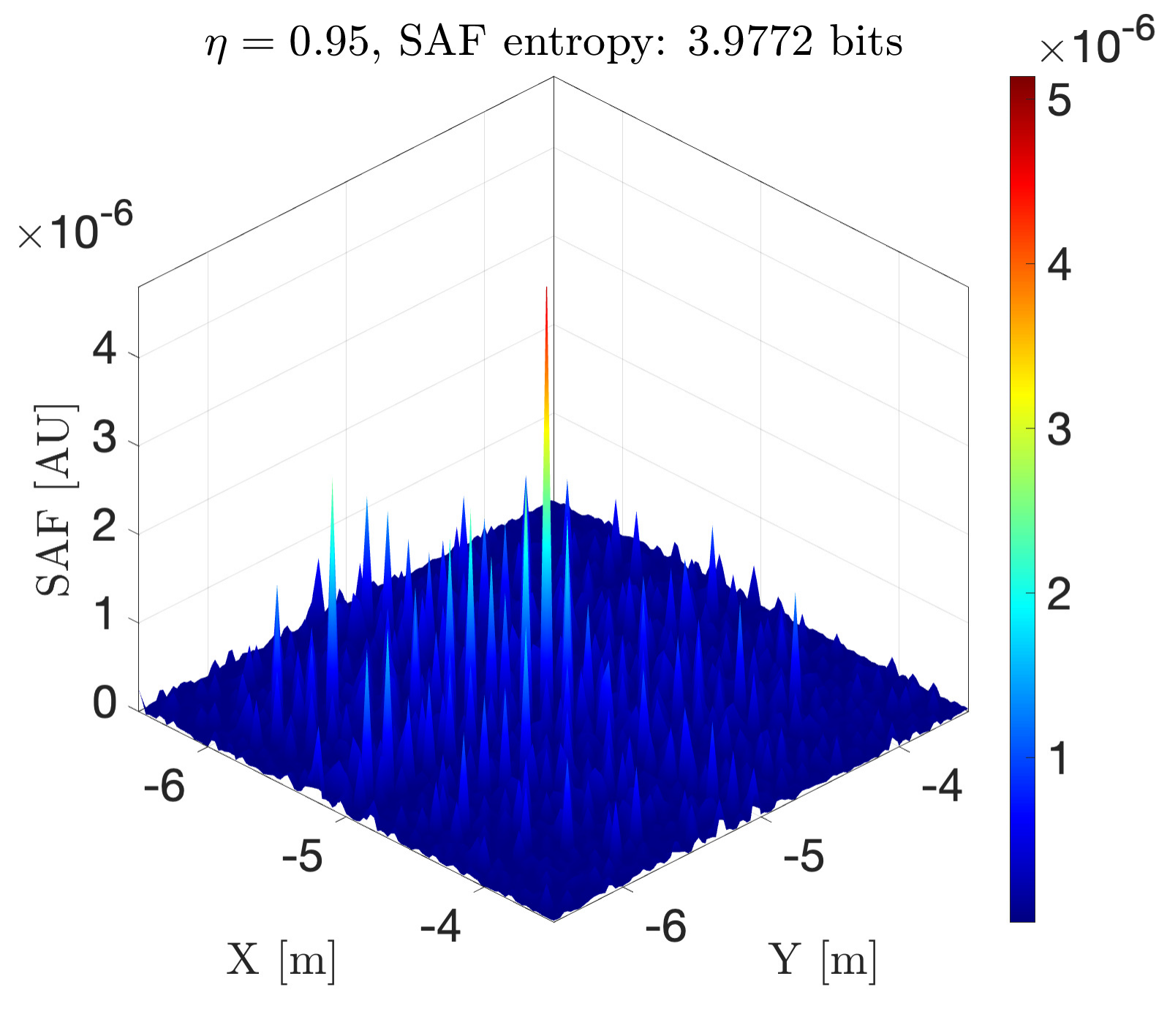}\label{subfig:entropy_vs_eta}}
\subfloat[][$\eta=0.99$]{\includegraphics[width=0.25\linewidth]{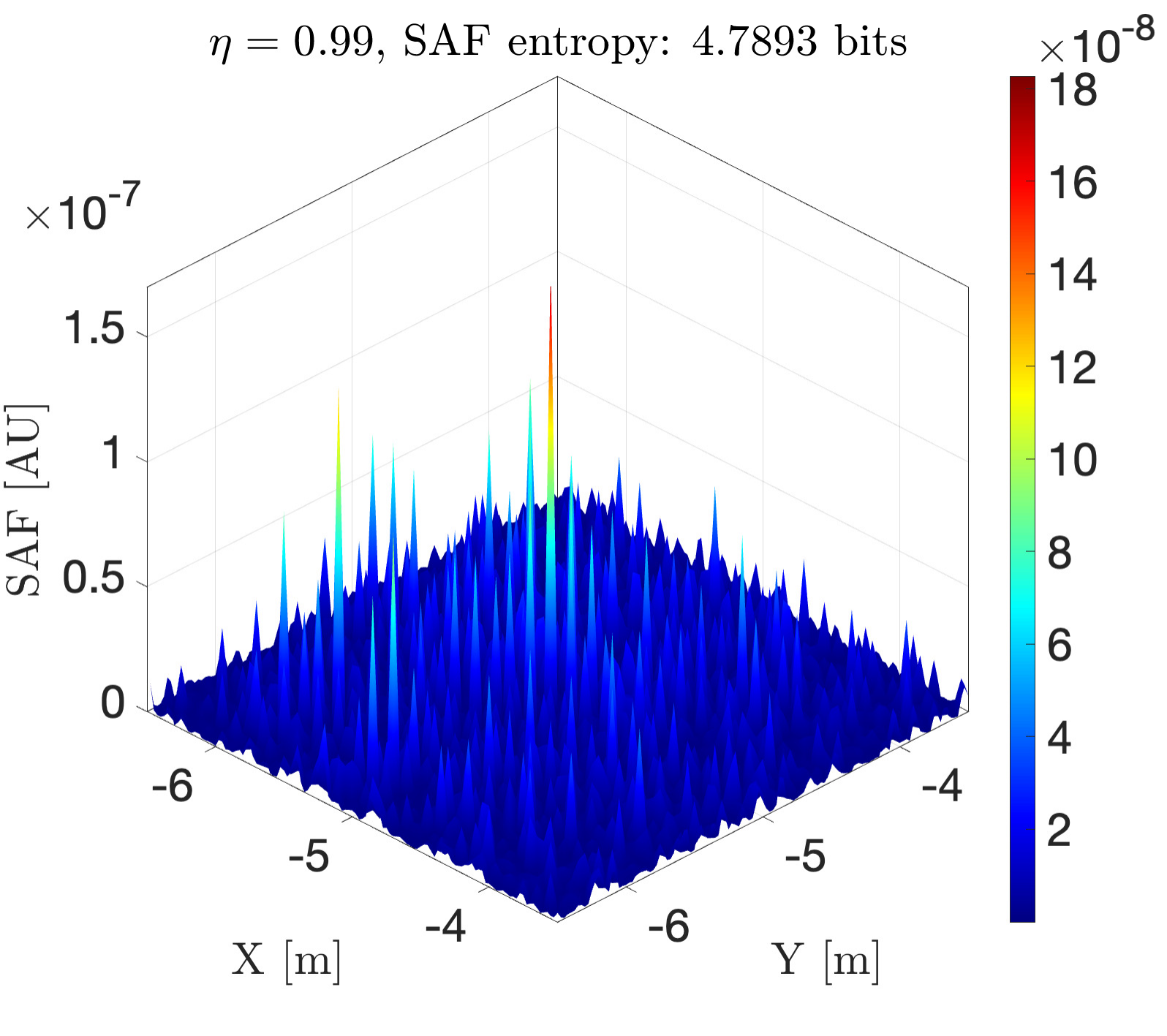}\label{}}
\caption{SAF for $N_{\rm rx}=1$, $M=128,K=1$ (a,b,c,d) and $N_{\rm rx}=4$, $M=512,K=4$ (e,f,g,h). The SAF should exhibit a single peak in $(-5,-5)$ and have nearly zero sidelobes. Due to disturbance (interference plus sparsity), sidelobes appear in the SAF. The lower-bound is the D-RN.}
\label{fig:SAF}
\end{figure*}

\vspace{-0.3cm}\subsection{Performance Analysis vs. Trade-Off Factor $\eta$}\label{subsect:performance_vs_eta}

The first set of results refers to $N=9$ APs. Fig. \ref{fig:KPI_vs_eta} shows the average SE (\ref{subfig:SE_VS_eta_Q=2_B=100MHz}), the average sensing SINR (\ref{subfig:SINR_VS_eta_Q=2_B=100MHz}) and the SAF entropy (\ref{subfig:ENTROPY_VS_eta_Q=2_B=100MHz}), for two values of $N_{\rm rx}\in\{1,4\}$. Black curves represent the D-MIMO upper bound on SE and the D-RN bounds on SINR and entropy. As expected, the SE grows with $\eta$, and MMSE outperforms MR precoding for $\eta \rightarrow 1$ ($\eta >0.3$), since the SE is dominated by the MUI. The SE with MMSE precoding almost attains the D-MIMO limit only for $N_{\rm rx}=1$ ($7\%$ capacity gap), while $N_{\rm rx}=4$ yields a $25\%$ capacity gap, irrespective of the usage of the same amount of power over the D-ISAC network. At $\eta=0.95$, the sensing interference yields a capacity reduction of less than $0.5\%$ (dotted lines) w.r.t. the ideal case of perfect sensing signal cancellation (dashed lines). Therefore, the additional computational burden at the UE for sensing cancellation is not practically needed, justifying our initial assumption. 

Concerning sensing performance, the average sensing SINR in Fig. \ref{subfig:SINR_VS_eta_Q=2_B=100MHz} is shown for two resource budgets, $M=128, K=1$ (blue and red lines) and $M=512,K=4$ (green and yellow lines). As expected, the SINR decreases with $\eta$, as more power is used for communication and less for sensing. For $\eta> 0.9$, the SINR exhibits a drastic decrease, as a result of a communication-oriented precoding and small emitted power. Notice that the proposed sensing waveform design in Section \ref{subsect:waveform_design} allows gaining up to $20$ dB at $\eta\rightarrow 0$ (in sensing-centric ISAC operation) and $M=512,K=4$, i.e., where the sensing interference dominates over the communication disturbance, and it has a significant effect up to $\eta = 0.8$ with the selected setup. As a direct figure of metric for imaging, SAF entropy is reported in Fig. \ref{subfig:ENTROPY_VS_eta_Q=2_B=100MHz}. The entropy increases with $\eta$, as expected, and decreases from $N_{\rm rx}=1$ to $N_{\rm rx}=4$ irrespective of $M$ (blue vs. red and yellow vs. green lines), as more bistatic AP pairs contribute to the final coherent image. As intuitive, the more Rx APs we allocate (up to the optimal value), the more the SE reduces. Therefore, for opportunistic imaging in D-MIMO networks, it is recommended to use less than $25\%$ of the available APs as Rx to guarantee an acceptable SE reduction. Notice that here entropy is not substantially affected by sensing interference, as the overall SINR is consistently $>10$ dB for $\eta \leq 0.8$. However, this is strongly dependent on the D-ISAC network, as discussed in the following Section \ref{subsect:performance_vs_config}. For $\eta>0.9$ the entropy rapidly increases, yielding a poor image quality, dominated by disturbance. As a means to reduce the entropy at fixed $N_{\rm rx}$ and $\eta$, more FT resources can be used, either increasing the number of subcarriers (for fixed bandwidth) $M$ and/or the number of OFDM symbols $K$. This allows better rejection of both sensing and communication interference, since the SINR increases according to \eqref{eq:sensing_SINR_extended} and the entropy decreases as shown in Fig. \ref{subfig:ENTROPY_VS_eta_Q=2_B=100MHz}. As a comparison, the D-RN (full multistatic case, no communication) is not practically affected by $M,K$, as it is only due to the D-ISAC topology.   
The SAFs corresponding to the results in Fig. \ref{subfig:ENTROPY_VS_eta_Q=2_B=100MHz} are depicted in Fig. \ref{fig:SAF}, for $\eta \in \{0, 0.95, 0.99\}$ and against the D-RN one. The ideal SAF should exhibit a single, narrow peak in the center, and zero (ore nearly-zero) sidelobes. The first row of Fig. \ref{fig:SAF} (a,b,c,d) shows the SAF for $N_{\rm rx}=1$ and $M=128,K=1$. The second row (e,f,g,h) is for $N_{\rm rx}=4$ and $M=512,K=4$. The SAF quality degrades with $\eta$ as predicted by the entropy, and at $\eta=0.99$, the image is completely corrupted, without a clear main lobe and dominated by disturbance from the communication signal. As expected, the D-RN shows the best SAF, characterized by the lesser sidelobes. By increasing $N_{\rm rx}$ and/or $M$ (Figs. \ref{fig:SAF} (e,f,g,h)) the disturbance is rejected and the SAF quality improves. In this case, even at $\eta=0.99$ (i.e., with approximately $98\%$ of the power allocated for communication), a $1.5$ bit reduction of the entropy makes the difference between a corrupted SAF (Figs. \ref{fig:SAF}(c)) and an acceptable one (Figs. \ref{fig:SAF}(d)), with a clear main lobe. Overall, the SAF is ultimately dependent the total number of APs (the D-ISAC network density) and the employed banwdith as discussed in the following. 

\begin{figure}[!b]
    \centering
    \subfloat[][SE]{\includegraphics[width=0.5\linewidth]{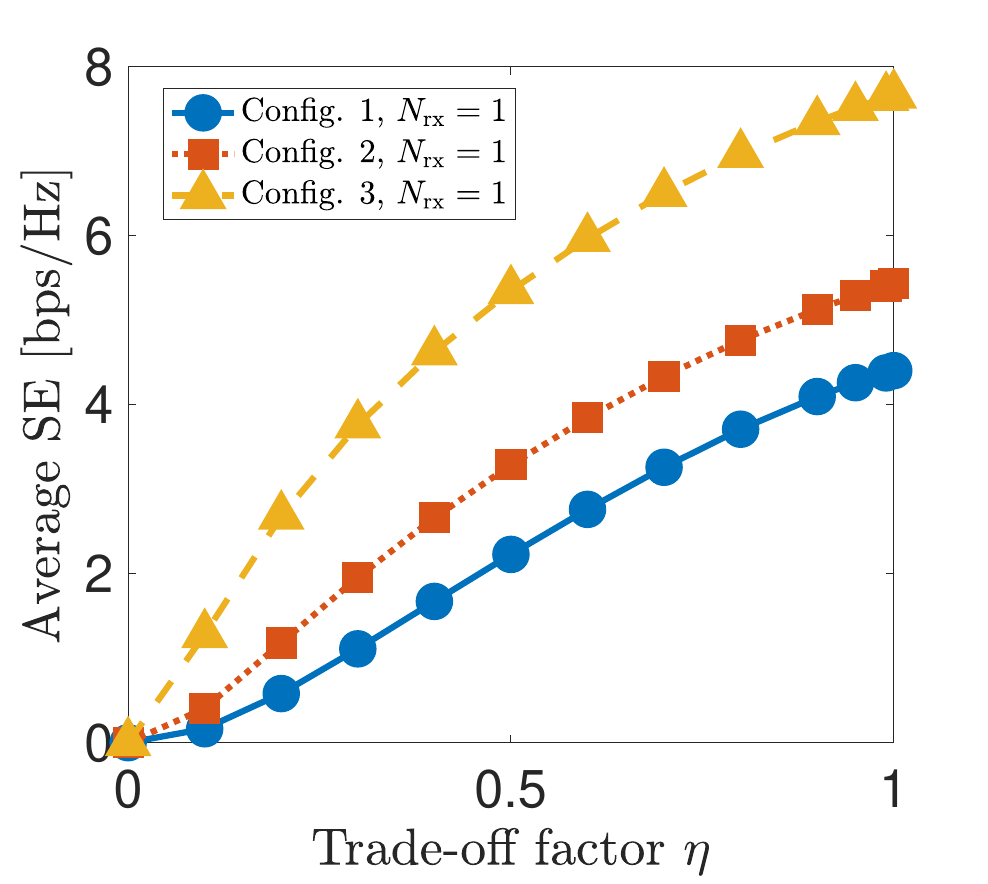} \label{subfig:SE_vs_config}} 
    \subfloat[][Entropy]{\includegraphics[width=0.5\linewidth]{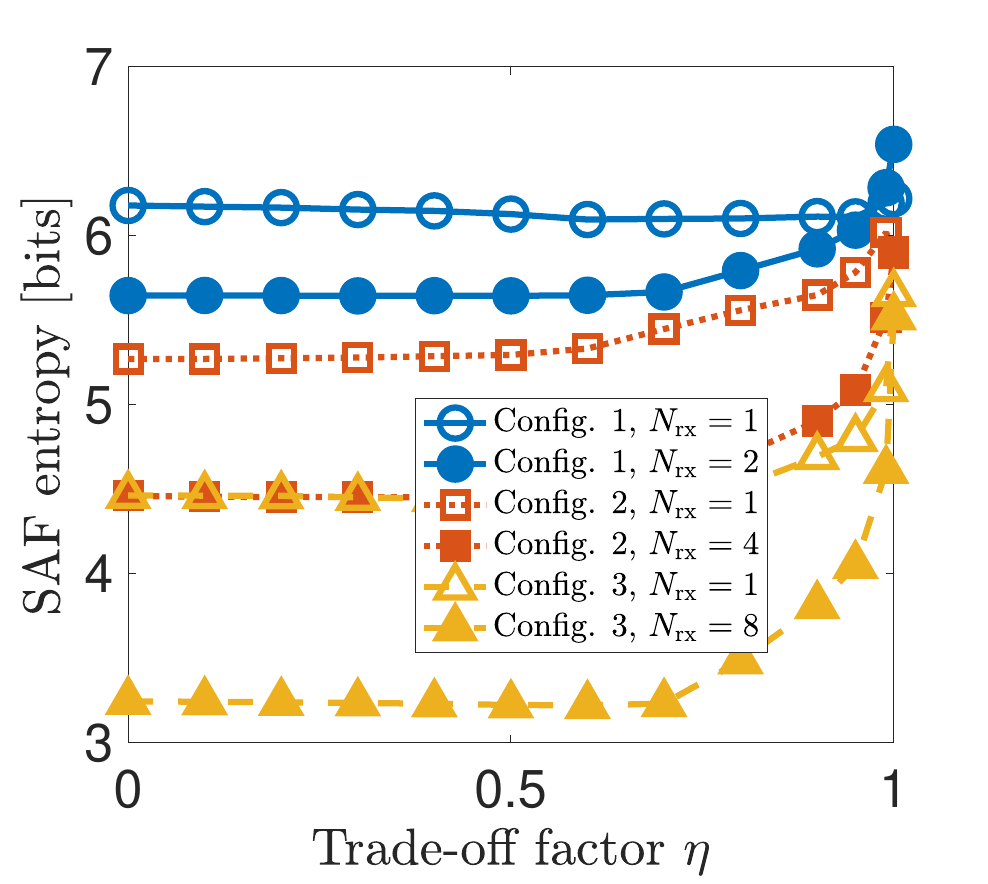}\label{subfig:entropy_vs_config}}
    \caption{SE (a) and SAF entropy (b) varying the trade-off factor $\eta$ for three D-ISAC configurations: Config. 1 (blue lines), Config. 2 (red lines), Config. 3 (yellow lines). }
    \label{fig:SE_entropy_vs_config}
\end{figure}

\begin{figure}
    \centering
    \includegraphics[width=0.95\linewidth]{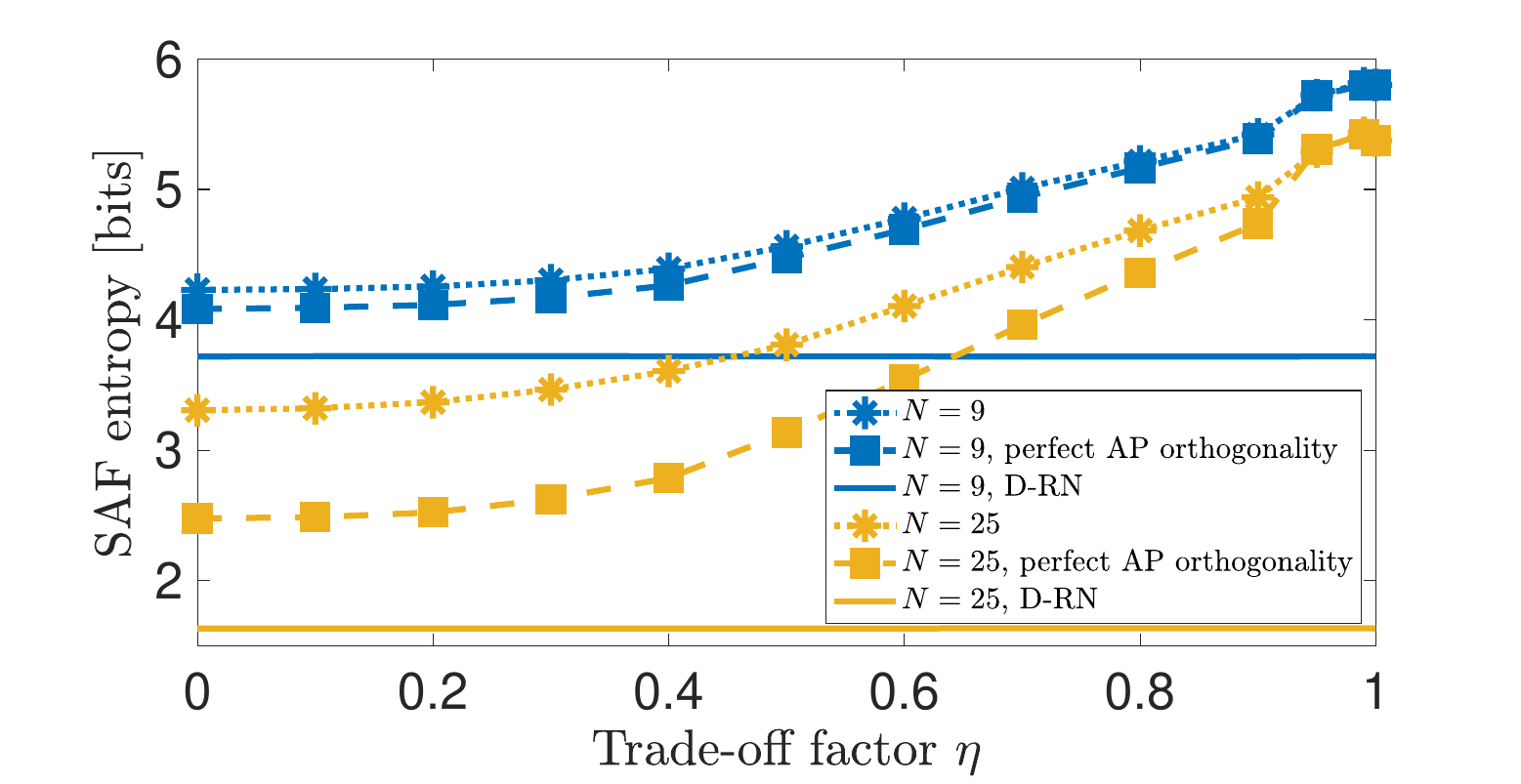}
    \caption{Comparison of SAF entropy for $N=9$ and $N=25$ for $L=2$ antennas and $N_{\rm rx}=1$ Rx AP, with and without waveform orthogonality according to Section \ref{subsect:waveform_design}.}
    \label{fig:Orthogonality_effect}
\end{figure}

\vspace{-0.3cm}\subsection{Performance Analysis vs. Number of APs $N$ and Number of Antennas $L$}\label{subsect:performance_vs_config}

The second set of results pertains to the comparison of different D-ISAC configurations for approximately fixed antenna budget, namely:
\begin{enumerate}
    \item \textbf{Config. 1}: $N=4$, $L=9$, $N_{\rm rx}\in\{1,2\}$
    \item \textbf{Config. 2}: $N=9$, $L=4$, $N_{\rm rx}\in\{1,4\}$
    \item \textbf{Config. 3}: $N=16$, $L=2$, $N_{\rm rx}\in\{1,8\}$
\end{enumerate}
The first represents a sparse network, the second a medium dense network and the last a comparably dense network. The results are shown in Fig. \ref{fig:SE_entropy_vs_config}. In Fig. \ref{subfig:SE_vs_config}, the SE vs. trade-off factor $\eta$ for $N_{\rm rx}=1$ Rx AP for MMSE precoding, and $M=512,K=4$. As expected, Config. 1 yields the lower SE due to the network sparsity and higher path-loss (on average) w.r.t. other configurations. Increasing the density of the network (reducing the number of antennas per AP) the average SE over the coverage area keeps increasing, as in conventional cell-free D-MIMO networks. Config. 3 shows the best result out of the tested configurations. Interestingly, increasing the spatial density of D-ISAC APs also yields better image quality. Fig. \ref{subfig:entropy_vs_config} illustrates the trend of the SAF entropy as function of $\eta$ for Config. 1,2,3. The entropy increases with $\eta$ but decreases with the network density (for fixed antenna budget), motivating the usage of dense D-ISAC networks made up of low-end nodes rather than massive nodes. The motivation for this behavior lies in that increasing the number of multistatic AP pairs leads to a densely sampled spatial spectrum (spectral coverage), and the SAF improves.

It is worth noticing the benefits of the proposed waveform design method according to extended orthogonality principle in dense D-ISAC networks: Fig. \ref{fig:Orthogonality_effect} compares the SAF entropy for $N=9$ and $N=25$ APs, for $L=2$ antennas and $N_{\rm rx}=1$ Rx AP, with and without waveform orthogonality according to Section \ref{subsect:waveform_design}. For non-orthogonal signaling, we opt for pseudo-random sequences at each Tx AP. The entropy gap for $N=25$ amounts to roughly 1 bit, which we have shown to be significant in Figs \ref{subfig:ENTROPY_VS_eta_Q=2_B=100MHz} and \ref{fig:SAF} (d and h). This encourages the use of the proposed waveform design method in dense D-ISAC settings. However, notice that dense D-ISAC networks for fixed ROI size require more resources as a consequence of the progressive null-space reduction described in Section \ref{subsect:waveform_design}. As $N$ increases, the waveform design defined by \eqref{eq:sensing_waveform_gen} not only needs more resources ($M$), but also must deal with an increase in cardinality of the set of delays $\mathcal{D}$. 

    \begin{figure}
    \centering
    \subfloat[][SINR]{ \includegraphics[width=0.5\linewidth]{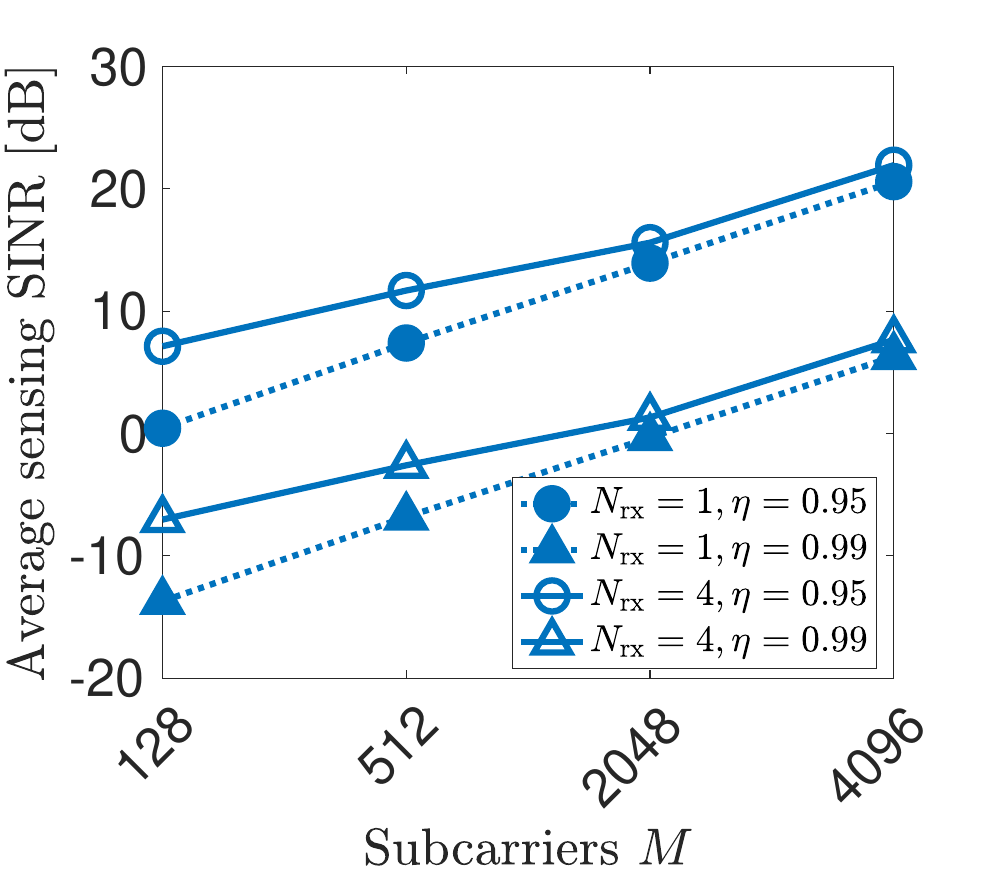}\label{subfig:SINR_vs_MK}}
    \subfloat[][Entropy]{ \includegraphics[width=0.5\linewidth]{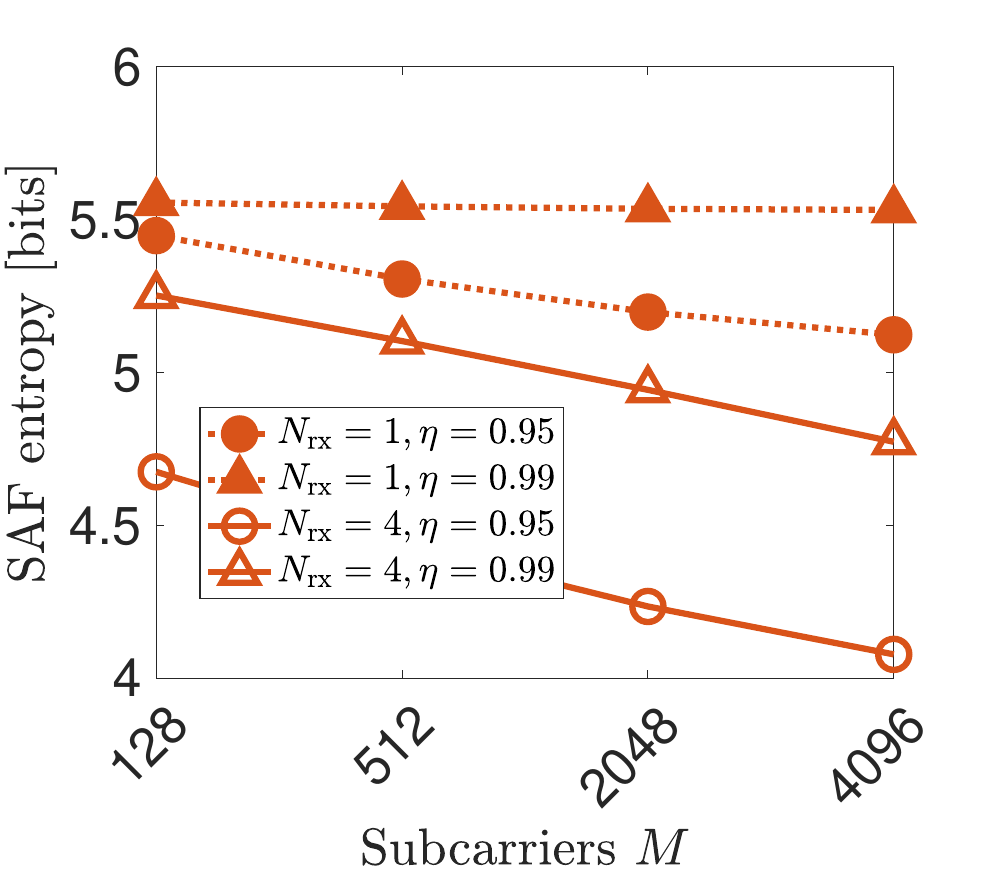}\label{subfig:entropy_vs_MK}}
    \caption{ Average Sensing SINR vs. SAF entropy varying the number of subcarriers $M$ for fixed bandwidth $B=100$ MHz.  }
    \label{fig:SINR_entropy_vs_MK}
\end{figure}\vspace{-0.3cm}

\vspace{-0.1cm}\subsection{Performance Analysis vs. Number of Subcarriers $M$}\label{subsect::performance_vs_subcarriers}

The third result details the effect of the FT resources $M$ on the sensing performance, for $N=9,L=4$ and fixed bandwidth $B=100$ MHz. We consider a communication-centric D-ISAC system, $\eta\in\{0.95,0.99\}$. Fig. \ref{fig:SINR_entropy_vs_MK} shows the SINR (Fig. \ref{subfig:SINR_vs_MK}) and the SAF entropy (Fig. \ref{subfig:entropy_vs_MK}) vs. the number of subcarriers $M\in\{128,512,2048,4096\}$ and $K=1$. By increasing $M$, the D-ISAC SINR increases, as expected. The entropy, instead, decreases with $M$, but for $N_{\rm rx}=1$ and $\eta=0.95$ is substantially limited by the D-ISAC network configuration, whose SAF is ruled by the comparably poor spectral coverage. Differently, for $\eta=0.99$ the decrease of the entropy is appreciable, especially for $N_{\rm rx}=4$, pushing for a small subcarrier spacing $\Delta f$ to have large values of $M$. Notice that decreasing $\Delta f$ for fixed $B$ implies longer OFDM symbol durations $T$, and in turn, longer DL bursts. As a general takeaway, the D-ISAC network has two options to improve imaging: either use more resources (but only for communication-centric operation,  $\eta \rightarrow 1$), tolerating longer DL bursts, or use short DL bursts and increase $N_{\rm rx}$ (valid for any value of $\eta$), tolerating a SE decrease.

 \begin{figure}
    \centering
    \subfloat[][$B=100$ MHz]{ \includegraphics[width=0.5\linewidth]{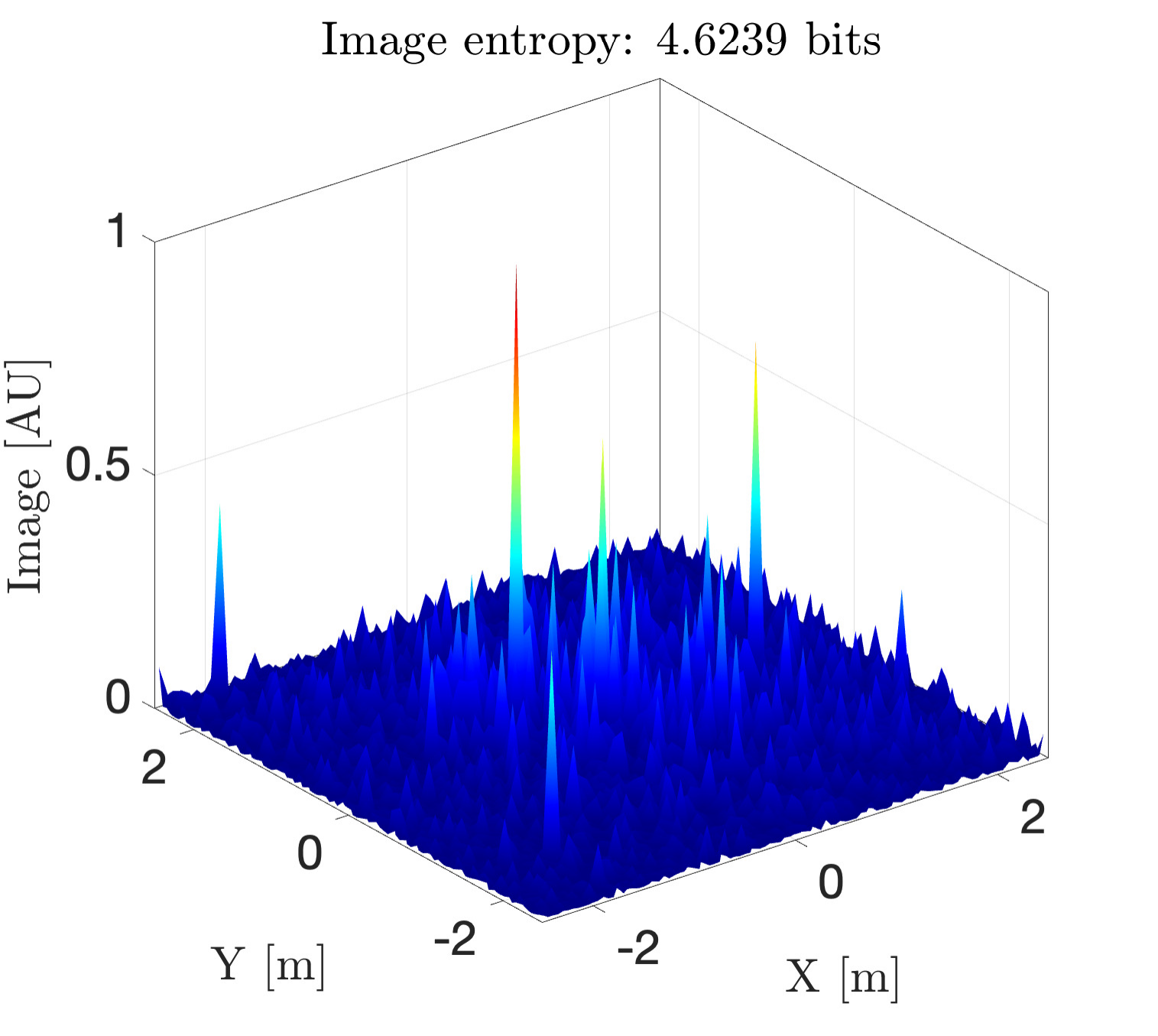}\label{subfig:Image:MT_B=100}}
    \subfloat[][$B=1$ GHz]{ \includegraphics[width=0.5\linewidth]{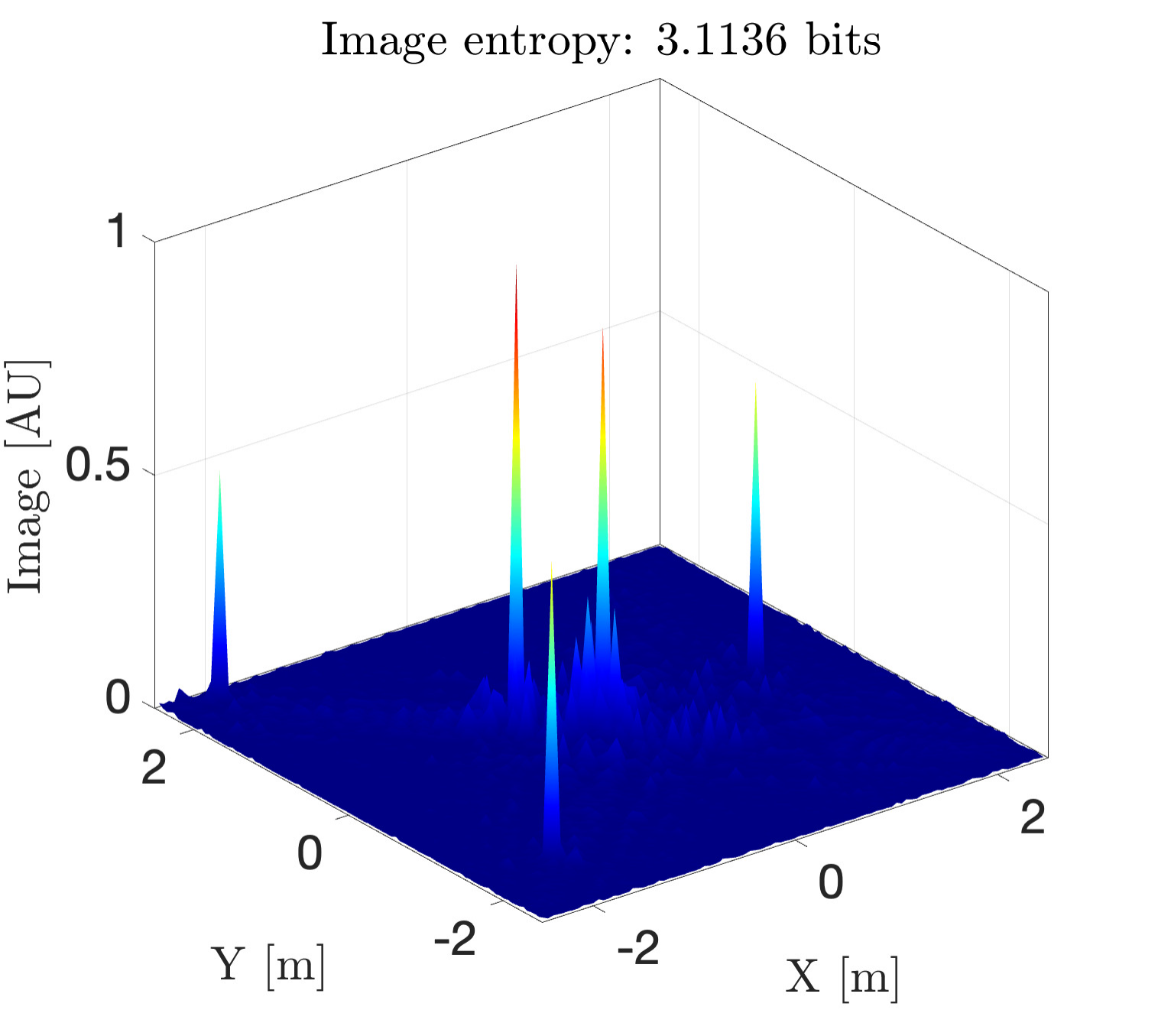}\label{subfig:Image:MT_B=400}}
    \caption{Multi-target coherent image for (a) $B=100$ MHz and (b) $B=1$ GHz.}
    \label{eq:image_vs_bandwidth}
\end{figure}

\vspace{-0.3cm}\subsection{Example of Multi-Target Image and Effect of Bandwidth $B$}\label{subsect:image_vs_B} 

The last result shows an example of a multi-target image, reported in Fig. \ref{eq:image_vs_bandwidth}, where $U=5$ point targets are deployed in the ROI, and the bandwidth changes from $B=100$ MHz to $B=1$ GHz. As expected, the bandwidth has a definite impact on the image quality: for $B=1$ GHz all the targets (the five peaks) are easily detectable, while just one of them appears with a significant peak in the $B=100$ MHz image. This is consistent with that found in recent literature~\cite{manzoni2024wavefield}: imaging is particularly effective for large fractional bandwidths $B/f_0$, as in the envisioned FR3 setup for 6G, where coherent aggregation over multiple subbands yields a compact spectral coverage and lower sidelobes. 

\vspace{-0.25cm}\section{Conclusion}\label{sect:conclusion}

This paper proposes a novel D-ISAC system to integrate multistatic coherent imaging into a D-MIMO network operating in DL. We first propose a space-time-frequency waveform design according to the extended orthogonality principle, to generate AP-specific orthogonal signals to be superposed to UE-specific OFDM signals, enabling imaging in the ROI with little or no effect at the UE side. Then, we outline an optimized selection strategy to cluster available APs into Tx and Rx and maximize imaging performance, owing to a half-duplex constraint. We study the communication-imaging trade-offs through extensive numerical simulations, revealing fundamental insights and design guidelines for a practical implementation. 
Future works on this research line would involve the consideration of extended (and thus non-isotropic) targets, the use of other metrics to evaluate imaging performance and the integration of our system design with existing imaging methods based on reconstruction by explicit inverse problem.

\bibliographystyle{IEEEtran}
\bibliography{bibliography.bib,Bibliography_TWC.bib,Bibliography_TWC2.bib,references.bib}

\end{document}